\documentclass[twocolumn,tighten]{aastex61}

\usepackage{graphicx}
\usepackage{ulem}
\usepackage{amsmath}
\usepackage{wrapfig}

\begin{document}
\shorttitle{Core collapse in GCs}
\shortauthors{Kremer et al.}

\title{How initial size governs core collapse in globular clusters}
\author[0000-0002-4086-3180]{Kyle Kremer}
\affil{ Department of Physics \& Astronomy, Northwestern University, Evanston, IL 60208, USA}
\affil{ Center for Interdisciplinary Exploration \& Research in Astrophysics (CIERA), Evanston, IL 60208, USA}
\author[0000-0002-3680-2684]{Sourav Chatterjee}
\affil{Tata Institute of Fundamental Research, Homi Bhabha Road, Mumbai 400005, India}
\affil{ Center for Interdisciplinary Exploration \& Research in Astrophysics (CIERA), Evanston, IL 60208, USA}
\author [0000-0001-9582-881X]{Claire S. Ye}
\affil{ Department of Physics \& Astronomy, Northwestern University, Evanston, IL 60208, USA}
\affil{ Center for Interdisciplinary Exploration \& Research in Astrophysics (CIERA), Evanston, IL 60208, USA}
\author{Carl L. Rodriguez}
\affil{MIT-Kavli Institute for Astrophysics and Space Research, Cambridge, MA 02139, USA}
\author[0000-0002-7132-418X]{Frederic A. Rasio}
\affil{ Department of Physics \& Astronomy, Northwestern University, Evanston, IL 60208, USA}
\affil{ Center for Interdisciplinary Exploration \& Research in Astrophysics (CIERA), Evanston, IL 60208, USA}

\begin{abstract}

Globular clusters (GCs) in the Milky Way exhibit a well-observed bimodal distribution in core radii separating the so-called ``core-collapsed" and ``non-core-collapsed" clusters. Here, we use our H\'{e}non-type Monte Carlo code, \texttt{CMC}, to explore initial cluster parameters that map into this bimodality. Remarkably, we find that by varying the initial size of clusters (specified in our initial conditions in terms of the initial virial radius, $r_v$) within a relatively narrow range consistent with the measured radii of young star clusters in the local universe ($r_v \approx 0.5-5$ pc), our models reproduce the variety of present-day cluster properties. Furthermore, we show that stellar-mass black holes (BHs) play an intimate role in this mapping from initial conditions to the present-day structural features of GCs. We identify ``best-fit" models for three GCs with known observed BH candidates, NGC 3201, M22, and M10, and show that these clusters harbor populations of $\sim 50-100$ stellar-mass BHs at present. As an alternative case, we also compare our models to the core-collapsed cluster NGC 6752 and show that this cluster likely contains few BHs at present.  Additionally, we explore the formation of BH binaries in GCs and demonstrate that these systems form naturally in our models in both detached and mass-transferring configurations with a variety of companion stellar types, including low-mass main sequence stars, white dwarfs, and sub-subgiants.

\end{abstract}
\keywords{globular clusters: general--stars: black holes--stars: kinematics and dynamics--methods: numerical}

\section{Introduction}
\label{sec:intro}

\subsection{Globular Cluster Evolution}
\label{sec:evolution}

The study of the evolution of dense star clusters is motivated by the application of these systems to a variety of areas in astrophysics. As high density environments, star clusters, in particular the old globular clusters (GCs), are expected to facilitate high rates of dynamical encounters, which can lead to the formation of various stellar exotica, including low-mass X-ray binaries, millisecond pulsars, blue stragglers, and cataclysmic variables. Observations of the spatial distribution of GCs in their host galaxies provide constraints on the formation and evolution of galaxies, making GCs valuable tools for extragalactic astronomy. Additionally, over the past several years, GCs have been shown to be efficient factories of the merging binary black hole (BH) systems that may be observed as gravitational-wave sources by LIGO, Virgo, and LISA \citep[e.g.,][]{Moody2009,Banerjee2010,Ziosi2014,Rodriguez2015,Rodriguez2016a,Hurley2016,Chatterjee2017a,Chatterjee2017b, Breivik2016,Askar2017, Kremer2018c}. This, in addition to the discovery of gravitational waves emitted from merging BH binaries by LIGO \citep{Abbott2016a,Abbott2016b,Abbott2016c,Abbott2016d,Abbott2016e, Abbott2017}, has sparked renewed interest in understanding the formation and evolution of BHs in GCs.

The old GCs observed in the Milky Way feature a clear bimodality in observed core radius \citep[e.g.,][]{Harris1996,McLaughlin2005}, separating the so-called ``core-collapsed'' clusters from their relatively puffy counterparts. Our understanding of the evolution of dense star clusters, and in particular, the dynamical processes that may lead to or prevent core-collapse has a long and varied history that has been guided by the complementary efforts of numerical simulations and observations over the past several decades (see, e.g., \citet{Heggie2003} for a thorough review).

Because star clusters are self-gravitating systems with negative heat capacities, dynamical perturbations in a cluster naturally lead to a flow of energy from the strongly self-gravitating core to the relatively sparse halo. The negative heat capacity means that the core becomes even hotter as the result of these perturbations, increasing the flow of energy to the halo in a runaway process that leads to core-contraction and ultimately collapse. 
This ``core-collapse" can be halted by an energy source in the core, which is expected to arise from binaries. For some time, these binaries were thought to exclusively form dynamically through three-body binary formation \citep[e.g.,][]{Heggie2003}; however, since the early 1990s, when primordial binary populations began to be observationally motivated, theoretical analyses have focused on studying properties of clusters with primordial binary populations as they pass through the so-called ``binary-burning'' phase, where the cluster core is supported against collapse by super-elastic dynamical scattering interactions of binary stars \citep[e.g.,][]{Vesperini1994,Fregeau2007,Chatterjee2013a}. 

Arguably, the most important recent shift in our understanding of how GCs evolve came from the observational and theoretical confirmation that GCs contain dynamically important populations of stellar-mass BHs up to the present time. 
Being the most massive objects in a GC, the BH population ``collapses" quickly and generates energy through dynamical binary formation, binary-burning, and dynamical ejections (see Section \ref{sec:BHs} for details and references). However, it is important to distinguish this BH collapse from the traditional observational definition of core-collapse in GCs. In particular, this BH collapse leaves little signature on the shape of the light profile of the GC which is sensitive only to the luminous stars \citep[e.g.,][]{Chatterjee2017a}. The only effect on the surface brightness profile is indirect: through strong dynamical encounters in the inner, BH-dominated region, BHs are frequently ejected to higher orbits in the cluster potential, leading to interactions with luminous stars in the outer parts of the cluster. Through these interactions, the BHs deposit energy into the GC's stellar bulk, leading to ``puffier" surface brightness profiles \citep[e.g.,][]{Mackey2007,Mackey2008,Kremer2018d}.

Thus, this BH collapse is very different from the formation of a cusp in the surface brightness profile which is the traditional observational definition of core-collapsed GCs. Most recently, several analyses have shown that only after the stellar-mass BH population is significantly depleted, can the surface brightness profile of a GC reach a traditional core-collapse architecture \citep[e.g.,][]{Mackey2008,Kremer2018d}. At this stage, in absence of a large number of BHs, the luminous binaries become the dominant source of energy at the GC's center.  In this study, we use the term ``core-collapsed'' to simply denote clusters (and cluster models) that are relatively centrally concentrated and have surface brightness profiles with prominent central cusps.

Observations of young massive clusters \citep[e.g.,][]{Holtzman1992,Whitmore1995,Miller1997,Bastian2005,Fall2005,Gieles2006,Scheepmaker2007,Scheepmaker2009,PortegiesZwart2010}, the expected progenitors of GCs, can provide insight into the various initial cluster properties that may determine the eventual outcome of the cluster, in particular, whether the GC has undergone core-collapse by the present day. Remarkably, observations indicate that although the masses of such young clusters can span several orders of magnitude, their sizes (e.g., core or half-light radii) span a relatively narrow range.

In this paper, we demonstrate that by exploring the small range in initial cluster size motivated by observations of young massive clusters, 
we produce a large spectrum of GC types at the present-day, ranging from core-collapsed clusters to puffy clusters with large core radii. We parameterize the cluster size in terms of the cluster virial radius, $r_v$, a theoretical quantity defined as

\begin{equation}
r_v = \frac{GM^2}{2\lvert U \rvert}
\end{equation}
where $M$ is the total cluster mass and $U$ is the total cluster potential energy, which can be calculated from the masses and positions of particles in our Monte Carlo calculation (Section \ref{sec:method}).

The initial relaxation timescale is directly related to the initial cluster size. Thus, the initial $r_v$ sets the dynamical clock of each cluster and controls how dynamically old a particular cluster is at a fixed physical time window, which, in turn, determines how close or far the cluster is from undergoing core-collapse. The half-mass relaxation time is given by

\begin{equation}
\label{eq:relaxation_time}
t_{\rm{rh}} = 0.138 \frac{M^{1/2}R_{\rm{h}}^{3/2}}{\langle m \rangle G^{1/2}\ln \Lambda}
\end{equation}
\citep[Equation 2-63 of][]{Spitzer1987}, where $M$ is the total cluster mass, $R_{\rm{h}}$ is the half-mass radius, $\langle m \rangle$ is the mean stellar mass, and $\ln \Lambda$ is the Coloumb logarithm where $\Lambda \simeq 0.4 N$, where $N$ is the total number of particles.

We demonstrate here that the evolution of stellar-mass BH populations in GCs, which is discussed at length in Section \ref{sec:BHs}, is intimately related to the contraction or expansion of the GC's core radius. In particular, since the initial $r_v$ of a cluster determines the initial relaxation timescale, the initial $r_v$ also controls how dynamically processed the BHs are at any given late physical time.

\subsection{Black Holes in Globular Clusters}
\label{sec:BHs}


Thousands of BHs are likely to form in GCs as the result of the evolution of massive stars. The number of these BHs that are \textit{retained} in GCs today is less certain. BHs are expected to be ejected from their host GCs through one of two primary mechanisms: ejection due to sufficiently large natal kicks or ejection via recoil as a result of strong dynamical encounters with other remaining BHs.

BH natal kicks, which are caused by asymmetric mass loss of supernova ejecta
are poorly constrained \citep[e.g.,][]{Belczynski2002,Belczynski2010,Repetto2012,Fryer2012,Mandel2016,Repetto2017}. If BH natal kicks are comparable in magnitude to the high speeds 
expected for the natal kicks of core-collapse NSs \citep[e.g.,][]{Hobbs2005}, the vast majority 
of BHs are likely to be ejected from GCs immediately upon formation because of the low escape speeds 
of typical GC cores. However, in the case of weaker BH natal kicks, a potentially large fraction of BHs may be retained post supernova.

The long-term retention of BHs that are not ejected promptly from natal kicks has long been a subject of debate. Until relatively recently, it was argued that BHs retained after formation would quickly mass-segregate and form a dense sub-cluster dynamically decoupled from the rest of the GC \citep[e.g.,][]{Spitzer1967,Kulkarni1993,Sigurdsson1993}. The BH members of this compact sub-cluster would then undergo strong dynamical encounters, ultimately ejecting all but a few BHs from the cluster on sub-Gyr timescales. However, more recently, several theoretical and computational analyses have demonstrated that this argument of rapid BH evaporation is not correct, and in fact, many BHs may be retained at present \citep[e.g.,][]{Merritt2004,Hurley2007,Mackey2007,Mackey2008,Morscher2015,Chatterjee2017a}.

The topic of retained BHs in GCs has been further motivated observationally. In the past decade, several stellar-mass BH candidates have been identified in both Galactic \citep{Strader2012,Chomiuk2013,Miller-Jones2014,Shishkovsky2018} and extragalactic \citep{Maccarone2007,Irwin2010} GCs. Most recently, the first stellar-mass BH to be identified through radial velocity measurements was found in the MW GC NGC 3201 \citep{Giesers2018}. The observations of these stellar-mass BH candidates suggest that at least some GCs do indeed retain populations of BHs at present and given that these host GCs do not show any particular trends in their observable properties, it appears that BH retention to present day may be common to most GCs. 

Several recent papers have used numerical simulations of GCs with large numbers of retained BHs to examine possible observational signatures that may indicate the presence of BH populations in GCs. \citet{Askar2018} predicted 29 MW GCs likely to have large BH subsystems, including two clusters considered here (M22 and NGC 3201) using a combination of numerical GC models and observations of MW GCs. \citet{Weatherford2017} demonstrated that a measure of mass segregation can be a robust observational tool to constrain unseen retained BH populations in GCs and predicted the number of BHs in three MW GCs (47 Tuc, M10, and M22) by correlating the size of BH populations with observational measurements of mass segregation.

Additionally, several previous analyses have used numerical simulations to model specific MW clusters known to harbor stellar-mass BH candidates. For example, \citet{Sippel2013} used N-body methods to model the MW GC M22 (known to contain \textit{two} stellar-mass BHs) and demonstrated that M22-like models can retain moderate numbers of BHs at late times. Shortly thereafter, \citet{Heggie2014} modeled M22 using Monte Carlo methods, and demonstrated similar results. More recently, \citet{Kremer2018d} used Monte Carlo methods to model NGC 3201 and showed that $\gtrsim 200$ BHs are necessary to produce models with observational features matching this cluster. \citet{Kremer2018d} showed that BHs are readily found in binaries with luminous companions (LCs) in BH-retaining clusters, although, \citet{Chatterjee2017a} and \citet{Kremer2018a} demonstrated that the presence of BH--LC binaries in a GC is uncorrelated with the total retained population of BHs. 


In general, the natal kick strengths determine the fraction of BHs retained immediately post formation. Subsequently, the cluster ejects BHs via dynamical processing including mass segregation and strong scattering over several relaxation times. In \citet{Kremer2018d}, we used the highly uncertain magnitudes of BH natal kicks to vary the number of BHs retained post supernova to ultimately control the retention fraction in GC models today. 
In this work, we vary the initial virial radius of the models to control the initial relaxation timescale to ultimately control how dynamically processed the BHs are at any given late physical time. 

We present a new grid of Monte Carlo GC models with varying initial virial radii and identify the models that best match three MW GCs in which stellar-mass BH candidates have been identified: NGC 3201 \citep{Giesers2018}, M22 \citep[two BH candidates; ][]{Strader2012}, and M10 \citep{Shishkovsky2018}. We demonstrate that the number of BHs retained in a GC has a significant effect upon the long-term evolution of the cluster. We show that both M22 and M10 likely contain $\sim 40-50$ stellar-mass BHs at present, in agreement with the predictions made be other recent studies, in particular \citet{Weatherford2017} and \citet{ArcaSedda2018}. In agreement with \citet{Kremer2018d}, we also show that NGC 3201 contains $>100$ BHs. Additionally, we show that accreting BH binaries similar to the observed systems are naturally produced in our models that are most similar to these three clusters.

We also compare our models to the core-collapsed MW GC NGC 6752, which has similar total mass to NGC 3201, M10, and M22. We demonstrate that NGC 6752 likely contains few BHs at present.

In Section \ref{sec:method}, we briefly describe our numerical techniques and discuss our grid of GC models. In Section \ref{sec:results} we show our results and discuss the best-fit models for the MW GCs considered in this study. In Section \ref{sec:BHMTB}, we explore the dynamical formation of accreting BH binaries in our models through several possible formation channels and discuss our results in the context of several of the observed accreting BH binaries identified to date. We conclude and discuss our results in Section \ref{sec:conclusion}.

\section{Method}
\label{sec:method}

\begin{deluxetable*}{c|c|c||c|cc|c|c|c}
\tabletypesize{\scriptsize}
\tablewidth{0pt}
\tablecaption{Initial and final cluster properties for all models\label{table:models}}
\tablehead{
	\colhead{Model} &
    \colhead{$r_v$} &
    \colhead{$t_{\rm{rh}}$} &
    \colhead{$M_{\rm{tot}}$}&
    \colhead{$r_c$}&
    \colhead{$r_h$}&
    \colhead{$N_{\rm{BH}}$}&
    \colhead{$N_{\rm{BH-LC}}$}&
    \colhead{$N_{\rm{BH-MTB}}$}\\
    \colhead{} &
    \colhead{(\rm{pc})} &
    \colhead{(\rm{Myr})} &
    \colhead{($10^5\,M_{\odot}$)}&
    \multicolumn{2}{c}{(\rm{pc})}&
    \colhead{}&
    \colhead{}&
    \colhead{}
}
\startdata
1 & 0.5 & 49 & 1.58 & 0.23 & 1.38 & 2 & 7 & 2\\
2 & 0.6 & 64 & 1.94 & 0.26 & 1.53 & 11 & 7 & 3\\
3 & 0.7 & 81 & 2.09 & 0.75 & 1.82 &16 & 4 & 1\\
4 & 0.8 & 99 & 2.17 & 0.90 & 2.22 & 28 & 6 & 1 \\
5 & 0.9 & 118 & 2.21 & 0.93 & 2.61 & 38 & 9 & 4\\
6 & 1.0 & 138 & 2.24 & 1.72 & 2.78 & 50 & 5 & 1\\
7 & 1.5 & 255 & 2.26 & 2.76 & 4.26 & 111 & 13 & 0\\
8 & 1.75 & 321 & 2.31 & 1.70 & 4.45 & 109 & 4 & 0\\
9 & 2 & 392 & 2.42 & 2.75 & 5.36 & 201 & 16 & 4\\
10 & 3 & 721 & 2.33 & 5.07 & 6.49 & 315 & 14 & 0\\
11 & 5 & 1552 & 2.38 & 9.90 & 11.7 & 614 & 18 & 8\\
\enddata
\tablecomments{Column 2 shows the initial virial radius, $r_v$, used for each model. Column 3 shows the initial half-mass relaxation time, $t_{\rm{rh}}$, given by Equation \ref{eq:relaxation_time}. Columns 4-6 show properties of each model at $t=12$ Gyr. Note that all models form approximately $1500$ BHs initially. Column 7 shows the number of BHs that are retained at $t=12$ Gyr. Column 8 shows the number of distinct BH--luminous companion (BH--LC) binaries and column 9 shows the number of distinct mass-transferring BH binaries (BH--MTBs); both columns 8 and 9 are based on snapshots in the range 10 Gyr $<t<$ 12 Gyr.}
\end{deluxetable*}

We use our \texttt{Cluster Monte Carlo} code (\texttt{CMC}) to model the evolution of GCs. \texttt{CMC} is a fully-parallelized code that uses H\'{e}non-style Monte Carlo methods to model the long-term evolution of GCs \citep[for a review, see][]{Henon1971a, Henon1971b,Joshi2000,Joshi2001, Fregeau2003, Umbreit2012, Pattabiraman2013, Chatterjee2010, Chatterjee2013a,Rodriguez2018}. \texttt{CMC} uses the stellar evolution packages \texttt{SSE} \citep{Hurley2000} and \texttt{BSE} \citep{Hurley2002} to model the evolution of single stars and binaries and uses the \texttt{Fewbody} package \citep{Fregeau2004,Fregeau2007} to model the evolution of three- and four-body encounters. \texttt{CMC} has been developed over the past decade-plus and has been shown to agree well with the results of $N$-body simulations of GCs. For a review of the most up-to-date modifications to \texttt{CMC}, including the incorporation of post-Newtonian terms into all few-body encounters, see \citet{Rodriguez2018}.

We fix various initial cluster parameters, including: total particle number, $N=8 \times 10^5$; King concentration parameter, $w_o = 5$; binary fraction, $f_b=5\%$; metallicity, $\rm{Z}=0.001$; and Galactocentric distance, $d=8$ kpc. The initial mass function for all stars and the initial period distribution for all binaries are chosen as in \citet{Kremer2018d}.

We adopt the prescription for stellar remnant formation described in \citet{Fryer2001} and \citet{Belczynski2002}. Natal kicks for core-collapse NSs are drawn from a Maxwellian with dispersion width $\sigma_{\rm{NS}} = 265 \rm{km\,s}^{-1}$ \citep{Hobbs2005}.  Unlike \citet{Kremer2018d}, we use a fixed prescription for BH natal kicks. We assume BHs are formed with fallback and calculate the BH natal kicks by sampling from the same kick distribution as the neutron stars, but with the BH kicks reduced in magnitude according to the fractional mass of fallback material \citep[see][for more details]{Morscher2015}.

We vary the initial virial radius, $r_v$, between $0.5-5$ pc, as described in Section \ref{sec:evolution}. 

Table \ref{table:models} includes a list of all GC models used in this study, including the initial values of $r_v$ and $t_{\rm{rh}}$ (columns 2 and 3, respectively) as well as various cluster properties at $t=12$ Gyr including total cluster mass (column 4), ``observed" (Section \ref{sec:obs}) core and half-light radii (columns 5 and 6, respectively), as well as total number of BHs at $t=12$ Gyr (column 7). Column 8 shows the total number of distinct BH--luminous companion binaries (BH--LCs) that appear in snapshots in the range 10 Gyr $<t<$ 12 Gyr. Column 9 shows the number of these late-time BH--LCs that are found in mass-transferring configurations. Note that because a single BH can undergo many exchange encounters from 10--12 Gyr, the same BH may appear in binaries with different stellar companions in different cluster snapshots. Hence, the total number of distinct BH--LCs and BH--MTBs that appear at late times (columns 8 and 9) can be greater than the total number of BHs found at the single $t=12$ Gyr snapshot (column 7).

\subsection{Calculating observational parameters}
\label{sec:obs}
In order to compare our GC models to observational features of MW GCs, we construct two-dimensional spatial resolutions of each model at various snapshots in time throughout the course of the evolution of the model assuming spherical symmetry. Given the uncertainty in ages of MW clusters, we consider all snapshots at times in the range 10 Gyr  $<t<$ 12 Gyr as equally valid representations of the present-day old GCs.

Using \texttt{SSE}, \texttt{CMC} calculates the bolometric luminosity and temperature of all stars versus time. We can determine the V-band luminosities by approximating each star as a blackbody and integrating the total luminosity in the V-band frequency range. From our randomly-generated two-dimensional cluster snapshots, we construct surface brightness profiles (SBPs) by dividing each cluster into 50 equally-spaced (in $\log$) radial bins and then calculate the total V-band luminosity of each radial bin by adding the contributions of all stars in the bin. We exclude from this calculation all stars with $L_{\star} > 15\, L_{\odot}$ to reduce the noise of a small number of bright stars. 

We estimate the observational half-light radius, $r_{\rm{hl}}$, of each model by finding the projected radius which contains half of the cluster's total light. We use the method described in \citet{Morscher2015} and \citet{Chatterjee2017a} to estimate the observational core radius, $r_c$. Note that the observed core radius is different than the theoretical (mass-density weighted) core radius (which we denote as $r_{\rm{c,\,theoretical}}$) traditionally used by theorists \citep{Casterano1985}. The distinction between $r_{\rm{c,\,theoretical}}$ and the observed $r_c$ is discussed further in Section \ref{sec:results}, particularly pertaining to Figure \ref{fig:rcrh}. Henceforth, unless otherwise noted, when using the term $r_c$, we refer to the core radius in the observational sense.

To calculate the velocity dispersion profile for each cluster, we implement the binning method of \citet{Zocchi2012}. As in \citet{Zocchi2012}, we include only giants in our velocity dispersion calculation and group the stars into radial bins of 25 stars each.

\subsection{Determining best-fit models}
\label{sec:bestfit}

We use a $\chi^2$ method to determine the goodness-of-fit of the models \citep[similar to that implemented in][]{Heggie2014} and identify the best-fit model(s) for each observed cluster. For each model, we consider all cluster time snapshots in the range $10-12$ Gyr. We also allow for uncertainty in the heliocentric distance to each cluster which shifts the SBPs by small amounts horizontally in either direction which, in turn, alters the fit to the observational data. We adopt distances from \citet{Harris1996}. The distances quoted in this catalog have no formal error bars, so we simply adopt distance errors of $10\%$. 

We calculate a single measure of the dispersion of error (rms) for each data point in the published SBPs and $\sigma_v$--profiles compared to the profiles calculated for each model snapshot. For each snapshot, we calculate a measure of goodness-of-fit, $\alpha = \chi_{\rm{SBP}}^2 + \chi_{\sigma_v}^2$, where $\chi_{\rm{SBP}}^2$ and $\chi_{\sigma_v}^2$ are the sums of rms values for all data points in the model SBPs and $\sigma_v$--profiles, respectively, compared to the observed data.

The data points in the observed SBPs and $\sigma_v$--profiles do not necessarily coincide with the radial binning used to construct the profiles for our \texttt{CMC} models. Thus, to calculate the rms value for each data point, we interpolate our model profiles to align with the radial locations of the points in the different observed profiles. In some cases, the model profiles extend to $r$-values where no observed data are available (in particular, for observed $\sigma_v$--profiles, which typically do not contain data points below $\approx 10$ arcsec). In this case, we construct the $\alpha$ statistic only over the available range of the observed profiles. To reflect the uncertainty on the heliocentric distances for the observed clusters, we repeat this process for three different distances, $d_{\rm{Harris}}$, the distance quoted in \citet{Harris1996}, as well as $d_{\rm{Harris}} \pm 0.1 d_{\rm{Harris}}$, calculating the statistic $\alpha$ corresponding to each choice of distance.

Using this scheme, we identify the 10 model snapshots with the lowest values of the statistic $\alpha$ as our best representations of each respective cluster. We use these minimum-$\alpha$ models to predict the total numbers of BHs in each cluster.   For Figures \ref{fig:NGC3201}, \ref{fig:m10}, \ref{fig:m22}, and \ref{fig:NGC6752} in Section \ref{sec:results}, we choose a single ``best-fit" model by eye from these minimum-$\alpha$ models that minimizes the stochasticity in the innermost regions ($r \lesssim 1$ arcsec), which are susceptible to uncertainty due to small $N$ in these regions, and show the $\sigma_v$--profile for these ``best-fit'' models in the lower panel of these four figures. When determining the predicted value of $N_{\rm{BH}}$ for a given cluster (as shown in Table \ref{table:bestfit}), all minimum-$\alpha$ models are considered.

\section{Results}
\label{sec:results}

\begin{figure}
\begin{center}
\includegraphics[width=\columnwidth]{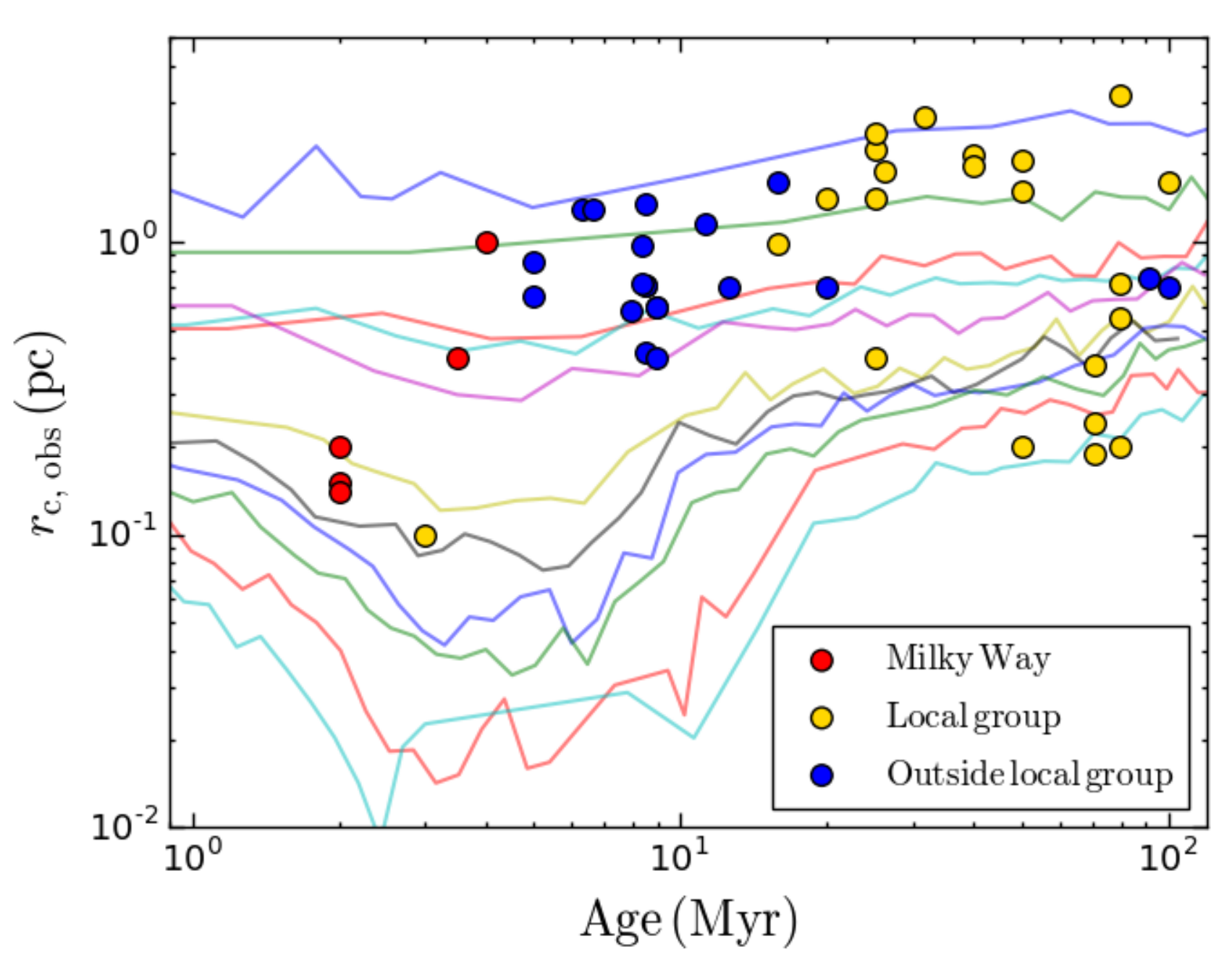}
\caption{\label{fig:rc_obs} Time evolution of ``observed'' core radii of all models for first 100 Myr of evolution. From bottom to top, the colored curves show models of increasing initial $r_v$. Filled scatter points mark observed core radii and ages of young massive clusters in the Milky Way (red), local group (yellow), and outside the local group (blue), taken from \citet{PortegiesZwart2010}.} 
\end{center}
\end{figure}

\begin{figure}
\begin{center}
\includegraphics[width=\columnwidth]{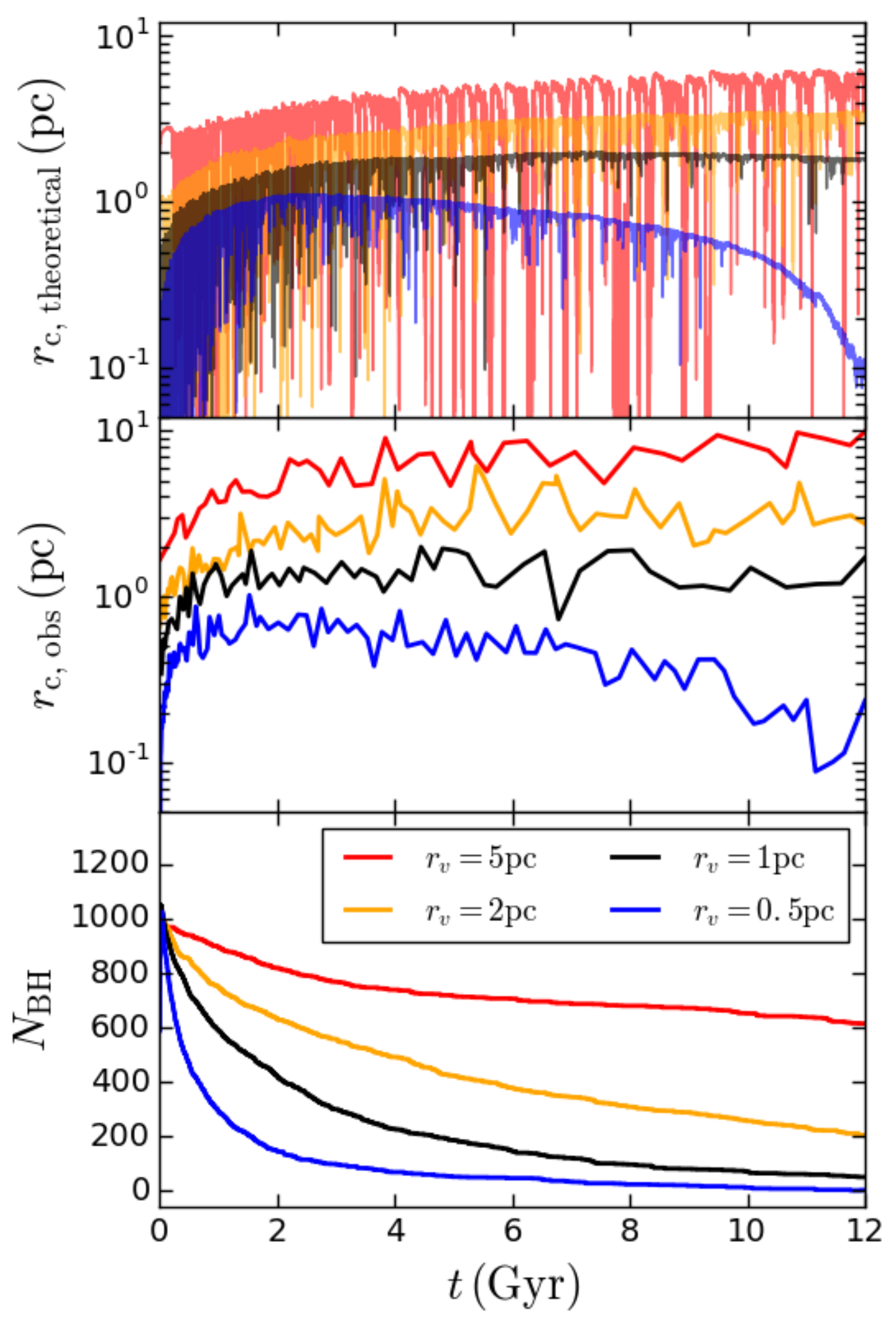}
\caption{\label{fig:rcrh} Theoretical core radius (top panel), observed core radius (middle panel), and total number of BHs (bottom panel) versus time for models with four different initial $r_v$: $r_v=5$pc (red curve), $r_v=2$ pc (orange curve), $r_v = 1$ pc (black), and $r_v = 0.5$ pc (blue).} 
\end{center}
\end{figure}

\begin{figure}
\begin{center}
\includegraphics[width=\columnwidth]{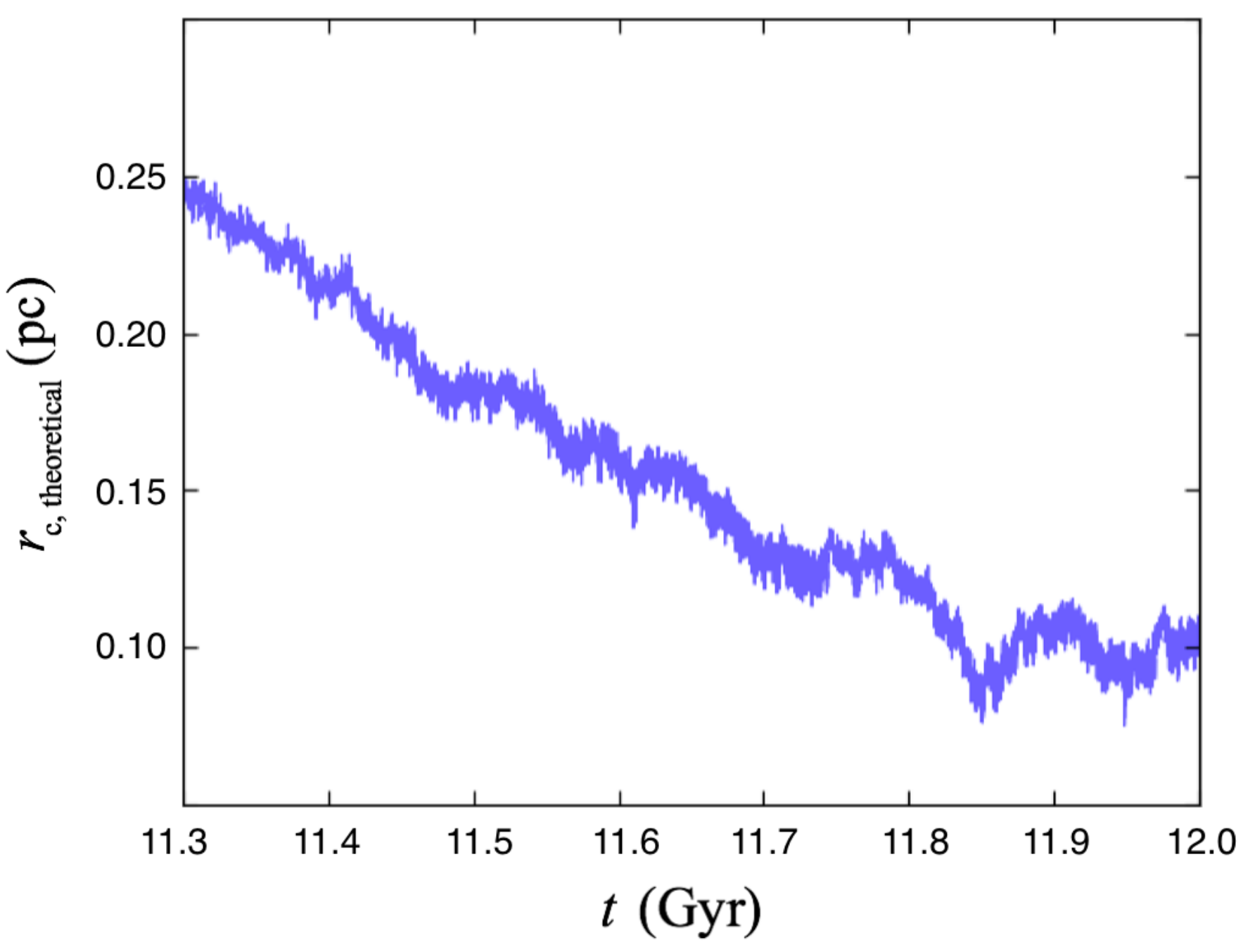}
\caption{\label{fig:binburning} Zoom-in on the $r_{\rm{c,\,theoretical}}$ for model 1 ($r_v=0.5$ pc; blue curve in Figure \ref{fig:rcrh}) from 11.3 - 12 Gyr. For $t \gtrsim 11.8$ Gyr, $r_{\rm{c,\,theoretical}}$ becomes flat, a sign of the onset of the traditional binary-burning phase \citep[e.g.,][]{Heggie2003} associated with a core-collapsed cluster.} 
\end{center}
\end{figure}

\begin{figure*}
\begin{center}
\includegraphics[width=0.95\linewidth]{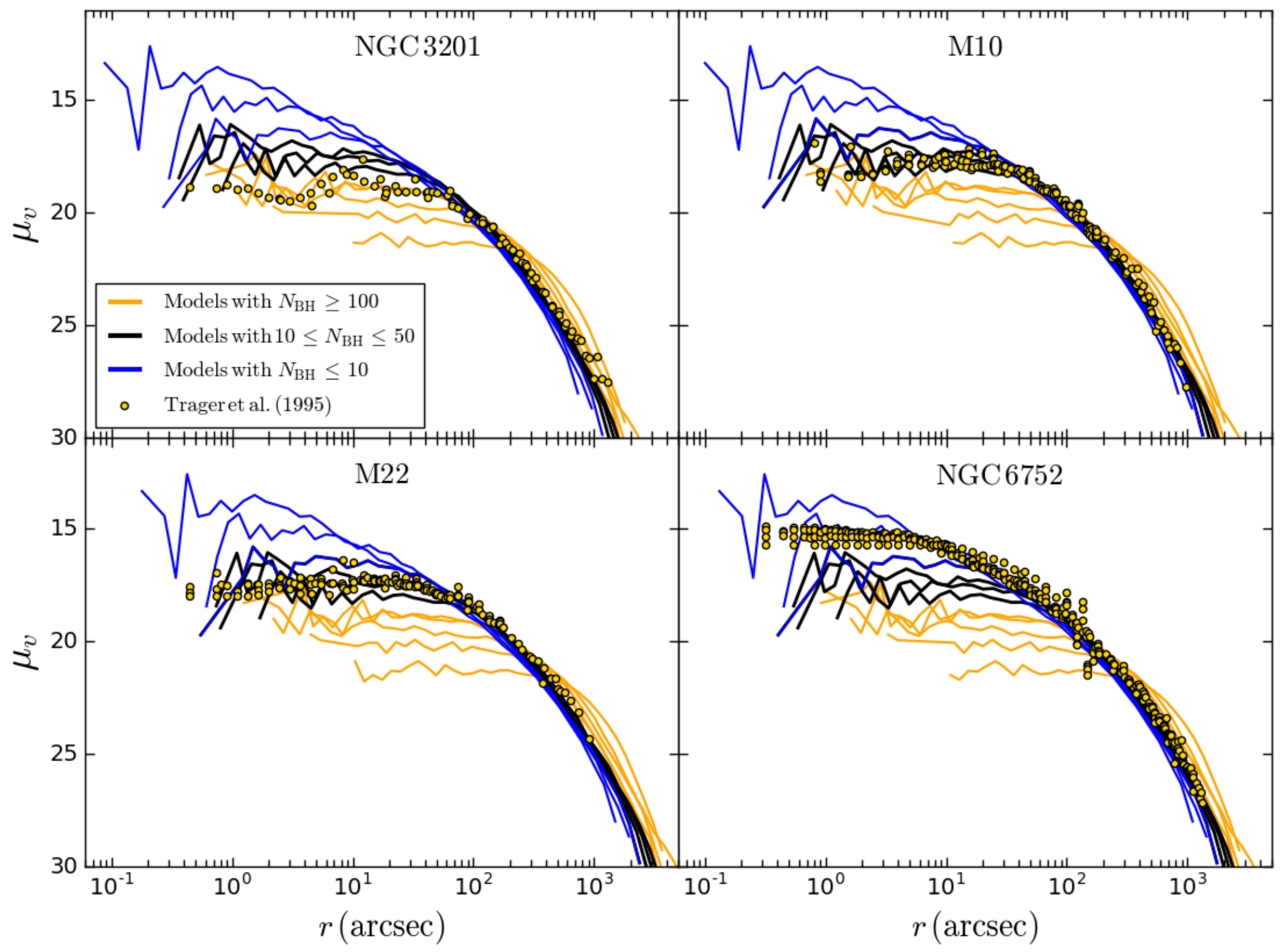}
\caption{\label{fig:SBP} Surface brightness profiles for all GC models listed in Table \ref{table:models} at $t=12$ Gyr. Orange curves denote BH-rich models ($N_{\rm{BH}} \geq 100$ at $t=12$ Gyr), blue curves denote BH-poor models ($N_{\rm{BH}} \lesssim 10$), and black curves denote models with intermediate number of BHs. The top-left panel shows our model SBPs compared to the observed SBP for NGC 3201 (gold circles), as studied in \citet{Kremer2018d}. The top-right panel shows the models compared to M10, bottom-left compares to M22, and bottom-right compares to the core-collapsed cluster NGC 6752. All observed SBPs are taken from \citet{Trager1995}. }
\end{center}
\end{figure*}

Figure \ref{fig:rc_obs} shows the time evolution of core radii (defined in the ``observational" sense; see Section \ref{sec:obs}) for the models listed in Table \ref{table:models}. From bottom to top, the colored curves show models of increasing initial virial radius. The scatter points mark observed core radii and ages \citep[taken from][]{PortegiesZwart2010}  for young massive clusters in the Milky Way (red), in the local group (yellow), and outside the local group (blue). Clearly, our selected range in initial virial radii ($0.5-5$ pc) effectively maps into the full range of observed core radii of young massive clusters in the local universe.

As Table \ref{table:models} shows, the number of BHs retained in the models at late times is directly related to the choice of initial $r_v$. Figure \ref{fig:rcrh} demonstrates this same result. Here we plot both the theoretical core radius (top panel), the observed core radius (middle panel), and the total number of BHs ($N_{\rm{BH}}$; bottom panel) versus time for four models: model 1 (blue curve; $r_v=0.5$ pc), which retains only 2 BHs at 12 Gyr; model 6 (black curve; $r_v=1$ pc), which retains 50 BHs; model 8 (orange curve; $r_v=2$ pc), which retains 201 BHs; and model 11 (red curve; $r_v=5$ pc), which retains 614 BHs. More precise best-fit models for each cluster are discussed in the following subsections, but at a basic level, the orange curve in Figure \ref{fig:rcrh} can be viewed as an NGC 3201-like cluster, the black curve can be viewed as an M10/M22-like cluster, and blue, as an NGC 6752-like (core-collapsed) cluster. 

In all four of these models, $\sim 1500$ BHs are formed initially, and $\sim 500$ of these are ejected promptly due to natal kicks. $N_{\rm{BH}}$ then gradually decreases over the course of the evolution of the cluster as BHs are slowly ejected through dynamical processing, as has been studied extensively and shown in previous analyses \citep[e.g.,][]{Morscher2015,Chatterjee2017a,Kremer2018d}.

As the bottom panel of Figure \ref{fig:rcrh} shows, the initial value of $r_v$ has a significant effect upon the way the BH population evolves over the lifetime of the cluster. Models with smaller initial $r_v$ (e.g., $r_v = 0.5$ pc) have shorter relaxation timescales (see Table \ref{table:models}), and therefore process their BHs faster. As a result, the evaporation timescale of BHs is shorter and fewer BHs are retained at late times. On the other side of the range, models with large initial $r_v$ (e.g., $r_v=5$ pc) have longer relaxation timescales. These models are less dynamically evolved by $t=12$ Gyr compared to their low-$r_v$ counterparts, and thus, retain larger fractions of BHs.

The top and middle panels of Figure \ref{fig:rcrh} show the time evolution of theoretical and observed core radii, respectively (see Section \ref{sec:obs}) for these four models. Because the distribution of stars (and thus the distribution of luminosities) in the central region of the cluster models varies from one time snapshot to the next, both the theoretical and core radii feature oscillations, as seen in the figure. However, unlike the observational $r_c$ shown in the middle panel, $r_{\rm{c,\,theoretical}}$ exhibits significantly more prominent variations on shorter time scales throughout its evolution. These sharp oscillations are a direct result of the formation of short lived cusps of the central-most BHs (see \citet{Morscher2015} for further discussion). We reiterate the discussion in Section \ref{sec:evolution} that these transient BH-collapse events are distinct from the observationally-defined core collapse which refers to the overall distribution of the cluster's luminous stars. Because the BHs do not contribute to the calculation of the observed $r_c$, these sharp cusps are absent from the curves shown in the middle panel.

The top and middle panels of Figure \ref{fig:rcrh} show that clusters with larger initial $r_v$ (and therefore, clusters which retain more BHs), exhibit larger core radii compared to clusters with smaller initial $r_v$ (and fewer BHs). Thus, high-$r_v$ models produce better representations of the relatively ``puffy'' GCs, such as NGC 3201, which has an observed core radius of 1.85 pc.


Figure \ref{fig:binburning} shows a zoom-in of model 1 (the blue curve of Figure \ref{fig:rcrh}) from 11-12 Gyr. For $t \gtrsim 11.8$ Gyr, $r_{\rm{c,\,theoretical}}$ becomes flat, a sign of the onset of the binary-burning phase \citep[e.g.,][]{Heggie2003}. With only a handful of BHs remaining at this time (see bottom panel of Figure \ref{fig:rcrh}), the non-BHs in this model enter the central regions. Densities for non-BHs increase to a point where super-elastic encounters involving luminous binaries become frequent enough to stall further core contraction and the cluster enters the traditional binary-burning phase involving luminous binaries. Equivalently, the SBP develops a clear cusp at the center, the traditional definition of a core-collapsed cluster.

Figure \ref{fig:SBP} shows the SBPs for all model clusters at 12 Gyr compared to the observed SBP \citep[from][]{Trager1995} for several Milky Way GCs: NGC 3201 (top left), M10 (top right), M22 (bottom left), and NGC 6752 (bottom right). Here the models are divided by color into three categories: BH-rich models (defined as $N_{\rm{BH}} > 100$; orange curves), BH-poor models (defined as $N_{\rm{BH}} < 10$; blue curves), and models with intermediate numbers of BHs (defined as $15 < N_{\rm{BH}} < 50$; black curves).

Just as Figure \ref{fig:rc_obs} demonstrates that our chose range of initial $r_v$ maps to observed features of young massive clusters, Figure \ref{fig:SBP} demonstrates the mapping to observed features of old GCs.

As Figure \ref{fig:SBP} clearly shows, BH-rich models (models 7--11 in Table \ref{table:models}) produce clusters most similar to NGC 3201 at late times, while BH-poor models (models 1--3) feature SBPs with prominent cusps at low $r$, representative of so-called ``core-collapsed'' MW GCs, such as NGC 6752.

 
Between the ``puffy'' and ``core-collapsed" extremes, we have models with intermediate number of BHs (models 4--6), that most accurately match the SBPs of M10 and M22. More precise ``best-fit models" for NGC 3201, M10, M22, and NGC 6752 are discussed in the following subsections.


Table \ref{table:bestfit}, shows the predicted number of stellar-mass BHs, BH-LC binaries, and BH-MTBs for NGC 3201, M10, M22, and NGC 6752, based on our best-fit models. Also included in the table is the range in initial cluster virial radii identified for the 10 minimum-$\alpha$ models for each cluster (gray curves shown in Figures \ref{fig:NGC3201}, \ref{fig:m10}, \ref{fig:m22}, and \ref{fig:NGC6752}). The best-fit models for each individual clusters are described in detail in the following subsections.

\begin{deluxetable}{l||c|c|c|c}
\tabletypesize{\scriptsize}
\tablewidth{0pt}
\tablecaption{Properties of best-fit models for various clusters \label{table:bestfit}}
\tablehead{
	\colhead{} &
    \colhead{NGC 3201} &
    \colhead{M10} &
    \colhead{M22}&
    \colhead{NGC 6752}
}
\startdata
$r_{v,0}$ (pc) & 1.75-2 & 0.7-0.9 & 0.8-0.9 & 0.5-0.7 \\
\hline
$N_{\rm{BH}}$ & $121 \pm 10$ & $39 \pm 9$& $40 \pm 9$ & $16 \pm 7$ \\
$N_{\rm{BH-LC}}$ & $2.5 \pm 0.5$ & $2.6 \pm 1.1$& $2.7 \pm 1.1$ & $2.7 \pm 1$ \\
$N_{\rm{BH-MTB}}$ & $1$ & $1.5 \pm 0.95$ & $1.5 \pm 1$ & $2.0 \pm 0.97$\\
\enddata
\tablecomments{The top row shows shows the range in initial virial radius, $r_{v,0}$, for the best-fit (minimum-$\alpha$) models for each cluster of interest. Row 2 shows the mean number of BHs at present (with $1\sigma$ uncertainties) calculated from these same best-fit models. Rows 3 and 4 show the mean number of BH--LC binaries and BH--MTBs for the best-fit models.}
\end{deluxetable}

\subsection{NGC 3201}

\begin{figure}
\begin{center}
\includegraphics[width=\columnwidth]{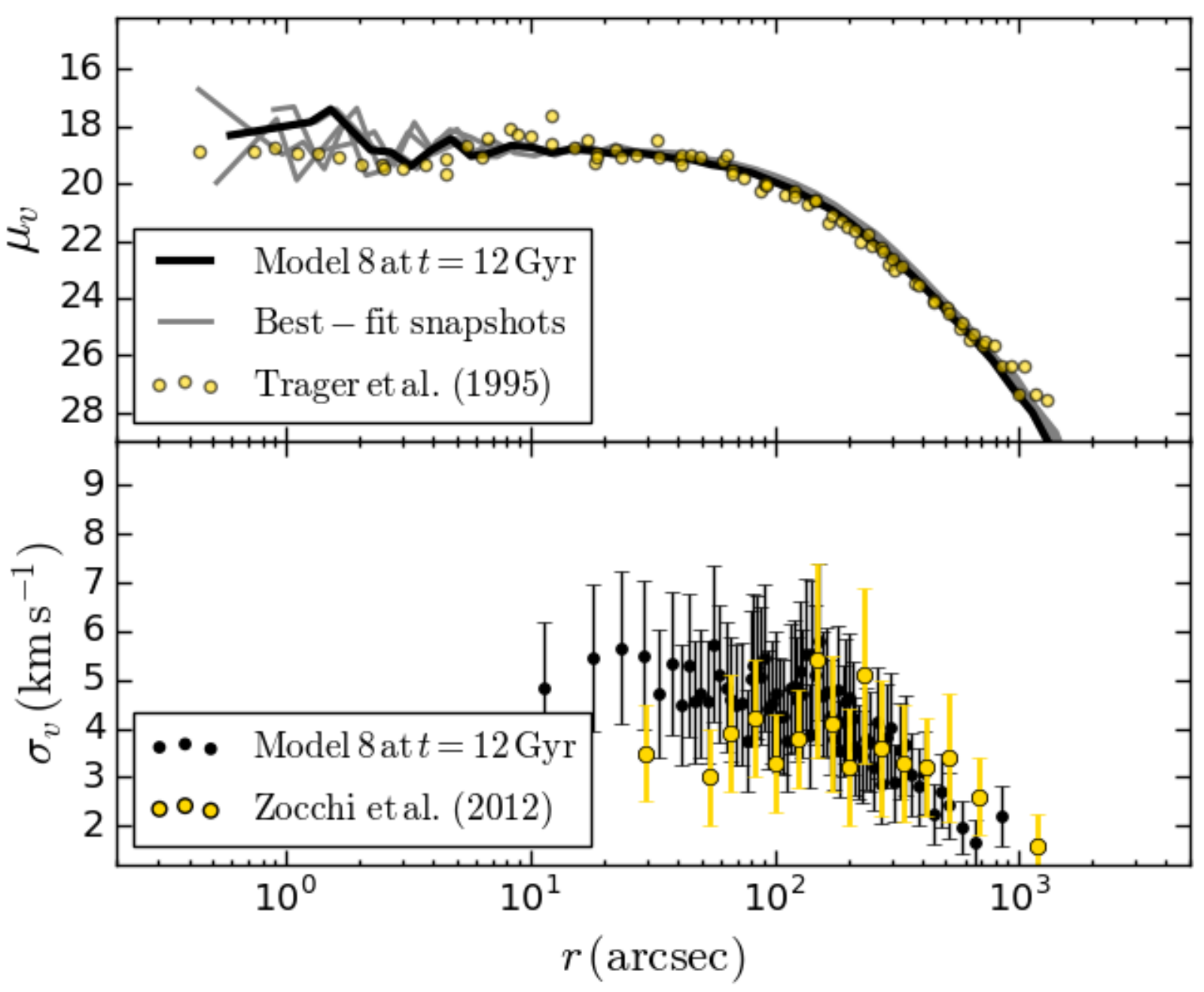}
\caption{\label{fig:NGC3201} Best-fit models for NGC 3201 compared to observations. Here, the bottom panel shows the $\sigma_v$-profile for model 8 at $t=12$ Gyr (shown as black curve in top panel) compared to the observed $\sigma_v$-profile from \citet{Zocchi2012}. We predict NGC 3201 contains $121 \pm 10$ BHs at present.} 
\end{center}
\end{figure}

As a follow-up to the results of \citet{Kremer2018d}, we first show our best-fit model for the cluster NGC 3201. The top panel of Figure \ref{fig:NGC3201} shows the SBPs of the best-fit models for NGC 3201 (the model snapshots with minimum $\alpha$, as discussed in Section \ref{sec:method}; gray curves) compared to the observational data of \citet{Trager1995}. The spread in gray curves about the observed SBP can be viewed as uncertainty on the models. Clearly, the model SBPs are most uncertain at small distances from the center ($r \lesssim 1$ arcsec), as expected due to the low $N$ in these regions.

We highlight one particular model (model 8 at $t=12$ Gyr; black curve), and plot the $\sigma_v$-profile for this model in the bottom panel of Figure \ref{fig:NGC3201} compared to the observed $\sigma_v$-profile of \citet{Zocchi2012} (shown here with $2\sigma$ errorbars).

Our best-fit model for NGC 3201 contains 109 BHs. From all of the best-fit models for NGC 3201 shown as gray curves in the top panel of Figure \ref{fig:NGC3201}, we predict NGC 3201 contains $120 \pm 10$ at present.

We note that this prediction is slightly less than that of \citet{Kremer2018d}, which showed NGC 3201 contains $\gtrsim 200$ BHs. This discrepancy results from our differing prescriptions for BH natal kicks. We intend to explore in more detail the effect of different natal kick prescriptions upon the evolution of GC BH systems in a future paper.

We also note that our predicted number of BHs is consistent with that of \citet{Askar2018}, who predicted that NGC 3201 contains $114\substack{+60 \\ -35}$ BHs. This agreement is satisfying given both studies used similar prescriptions for BH formation.

The present result also differs from \citet{Kremer2018d} in the number of BHs retained at birth. Using the fallback-based BH retention model (see Section \ref{sec:method}), roughly 1000 of the 1500 total BHs formed are retained in the cluster initially after natal kicks. This is in contrast to the best-fit model for NGC 3201 identified in \citet{Kremer2018d}, for which $\sigma_{\rm{BH}}/\sigma_{\rm{NS}}=0.04$, where a somewhat higher fraction of BHs (roughly 1400 out of 1500) are retained initially. The models in \citet{Kremer2018d} assumed initial virial radii of $r_v=1$ pc (with all other cluster parameters the same as here). As illustrated in Figure \ref{fig:rcrh}, clusters with initially smaller core radii will dynamically process their BHs more quickly, so it is not surprising that the $r_v=1$ pc model of \citet{Kremer2018d} needed to retain more BHs at birth (achieved by adopting smaller BH natal kicks) to achieve a similarly large population of BHs at late times compared to the $r_v=1.75$ pc model identified here.

\vspace{1cm}

\subsection{M10}

\begin{figure}
\begin{center}
\includegraphics[width=\columnwidth]{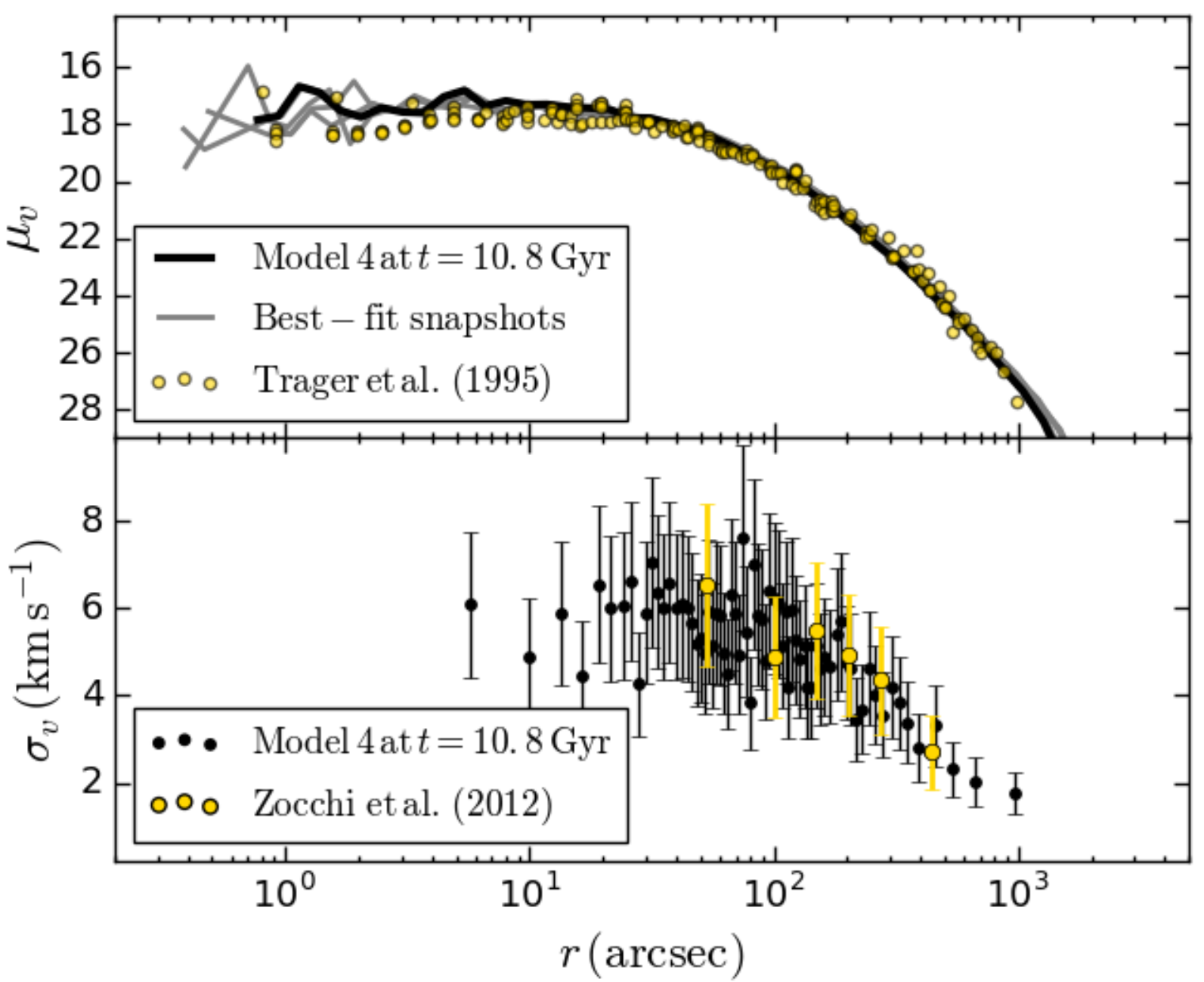}
\caption{\label{fig:m10} The top panel shows the SBP for the best-fit models for M10 (gray curves) compared to observations \citep{Trager1995}. The bottom panel shows the $\sigma_v$-profile for model 4 at $t=10.8$ Gyr (shown as black curve in top panel) compared to observations from \citet{Zocchi2012}. On the basis of these best-fit models, we predict that M10 contains $39 \pm 9$ BHs at present.} 
\end{center}
\end{figure}

Figure \ref{fig:m10} is analogous to Figure \ref{fig:NGC3201} but for the model snapshots that most accurately match the observed SBP and $\sigma_v$-profile of M10. The black curve in the top panel marks the SBP for model 4 at $t=10.8$ Gyr, the best-fit model for M10. At this time, this model contains 33 BHs, one of which is found in a BH--LC binary. Although this particular time snapshot does not contain an accreting BH--LC, other cluster snapshots in the late time range of 10--12 Gyr do contain accreting BH binaries (see Table \ref{table:models}).

From all of the best-fit models for M10 shown as gray curves in the top panel of Figure \ref{fig:m10}, we predict M10 retains $39 \pm 9$ BHs at present.

By correlating the size of BH populations with observational measurements of mass segregation and using cluster models also developed using \texttt{CMC}, \citet{Weatherford2017} predict that M10 contains up to 38 BHs, with a mode at 24, consistent with our result. 

\subsection{M22}
\label{sec:m22}

\begin{figure}
\begin{center}
\includegraphics[width=\columnwidth]{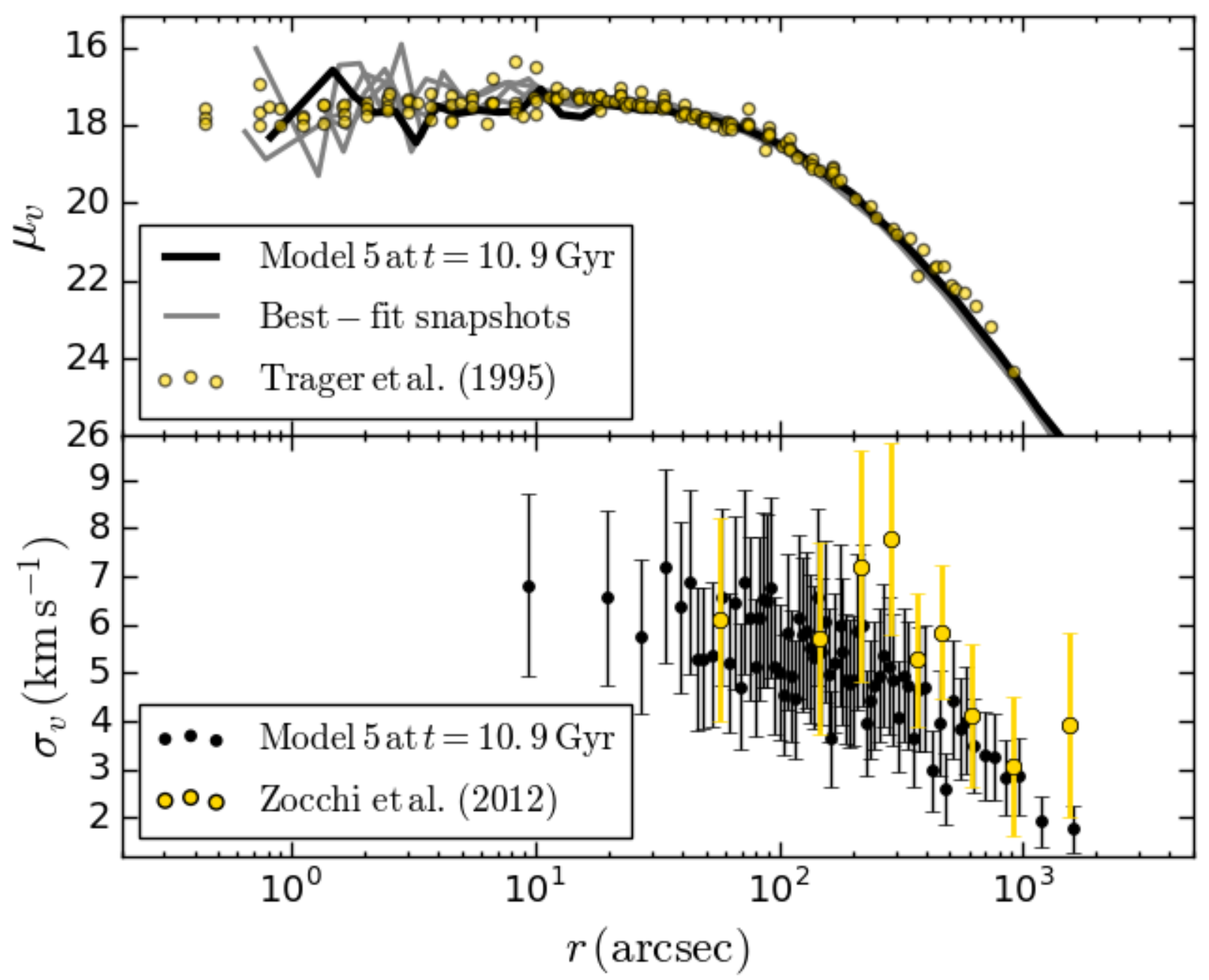}
\caption{\label{fig:m22} Same as Figure \ref{fig:m10}, but for M22. Here, the bottom panel shows the $\sigma_v$-profile for model 5 at $t=10.9$ Gyr (shown as black curve in top panel). We predict that M22 contains $40 \pm 9$ BHs at present.} 
\end{center}
\end{figure}

Figure \ref{fig:m22} shows the models that most accurately match the observed SBP and $\sigma_v$-profile of M22. The black curve in the top panel marks the SBP for model 5 at $t=10.9$ Gyr, the best-fit model for M22. At this snapshot in time, this model contains 49 BHs, four of which are found BH--LC binaries. Two of these four BH--LC binaries are found in mass-transferring configurations.

From all of the best-fit models for M22 shown as gray curves in the top panel of Figure \ref{fig:m22}, we predict M22 has $40 \pm 9$ BHs at present.

Several previous analyses have studied the BH population in M22, and drawn similar conclusions to those drawn here. On the basis of the two accreting stellar-mass BHs in M22 and on the expected fraction of BHs that will be found in accreting systems, \citet{Strader2012} argued that M22 likely contains $\sim 5-100$ stellar-mass BHs. Soon thereafter, \citet{Sippel2013} modeled M22 using direct $N$-body methods and imposed an initial BH retention fraction of 10\%. This analysis found that, for a model slightly less massive than M22 at present, 16 BHs were retained at $t=12$ Gyr. \citet{Heggie2014} used Monte Carlo methods similar to those considered in this study to model M22 and predicted $\sim 40$ BHs are likely retained at present. Most recently, using a combination of Monte Carlo GC models and observations of MW GCs, \citet{Askar2018} predicted that M22 retained $63\substack{+25 \\ -16}$ BHs at present, consistent with our predicted number.\citet{Weatherford2017} predict $49\substack{+50 \\ -34}$ BHs in M22, also consistent with our prediction.



\subsection{NGC 6752}

\begin{figure}
\begin{center}
\includegraphics[width=\columnwidth]{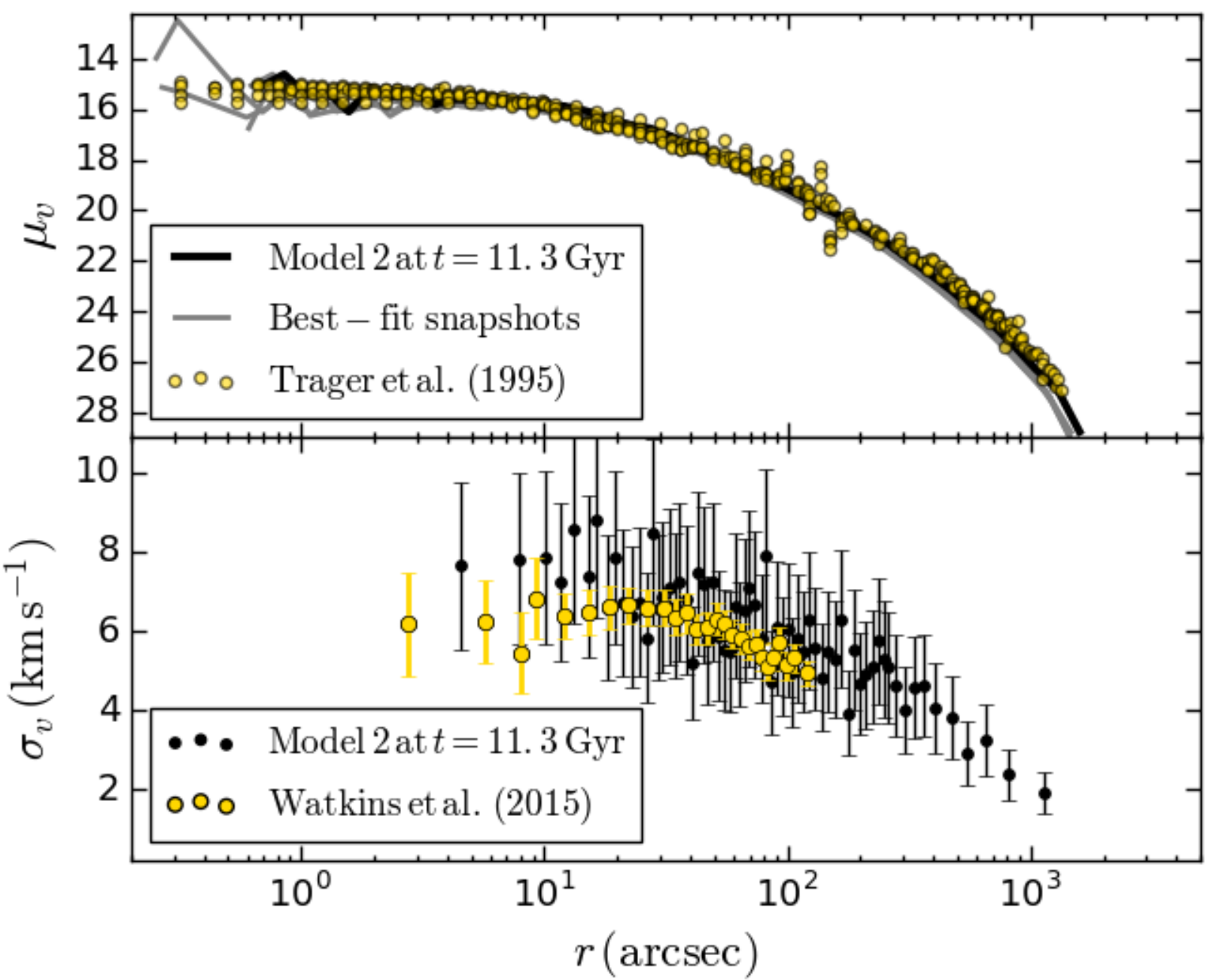}
\caption{\label{fig:NGC6752} Best-fit models for NGC 6752 compared to observations. Here, the bottom panel shows the $\sigma_v$-profile for model 2 at $t=11.3$ Gyr (shown as black curve in top panel) compared to the observed $\sigma_v$-profile from \citet{Watkins2015}. From our best-fit models, we predict that NGC 6752 contains $16 \pm 7$ BHs at present.} 
\end{center}
\end{figure}

Finally, Figure \ref{fig:NGC6752} shows the SBP and $\sigma_v$-profile for our best-fit models for NGC 6752. The black curve in the top panel marks model 2 at $t=11.3$ Gyr, which has 17 BHs. Since \citet{Zocchi2012} does not contain a $\sigma_v$-profle for NGC 6752, we compare to the $\sigma_v$-profile of \citet{Watkins2015}, shown in the bottom panel of the figure compared to the $\sigma_v$-profile of model 2. We predict NGC 6752 has $16 \pm 7$ BHs at present.

Unlike M10 and M22 (and NGC 3201), NGC 6752, does not contain an observed stellar-mass BH candidate. We include this cluster here to serve as an alternative case to the other three relatively ``puffy'' clusters. Observationally, NGC 6752 is classified as a core-collapsed cluster.

Note that the model representative of NGC 6752 also shows the well-known flattening of the theoretical core radius (e.g., see Figure \ref{fig:binburning}) representative of the binary-burning phase involving luminous stars commonly associated with core-collapse \citep[e.g.,][]{Chatterjee2013a}.


As shown in \citet{Kremer2018a} and \citet{Chatterjee2017a}, provided these core-collapsed clusters still retain at least a few BHs at late times, they are still just as likely to contain BH--LC binaries in both detached and mass-transferring configurations as the clusters with large populations of BHs. As shown in Table \ref{table:models}, those models which contain only a few BHs at $t=12$ Gyr (models 1--3) still produce up to 7 distinct BH--LC binaries and up to 3 accreting BH binaries.




\newpage

\section{Formation of Black Hole Binaries}
\label{sec:BHMTB}

\begin{figure}
\begin{center}
\includegraphics[width=\columnwidth]{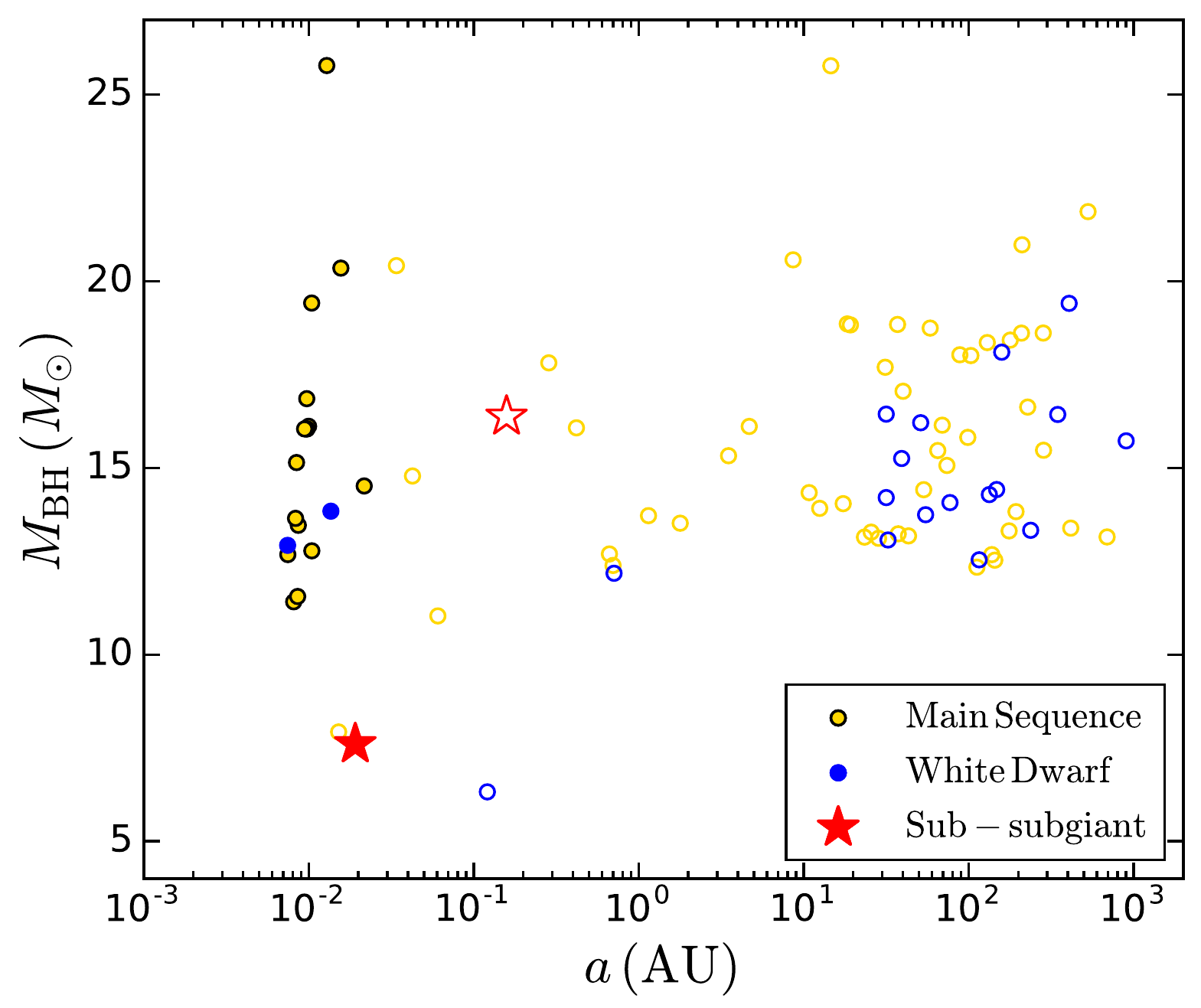}
\caption{\label{fig:scatter} BH mass versus semi-major axis for all BH--LC binaries found in our models at late times. Yellow represents systems with main sequence star companions, blue represent white dwarf companions, and red represent sub-subgiant companions. Open (filled) circles/stars indicate detached (mass-transferring) binaries.} 
\end{center}
\end{figure}

Stellar-mass BH candidates observed in binaries with luminous companions in clusters \citep[e.g.,][]{Strader2012,Chomiuk2013,Miller-Jones2014,Giesers2018,Shishkovsky2018} provide important constraints on BH populations in their host systems. The formation channels of these systems have been explored in the context of both globular clusters \citep[e.g.,][]{Sippel2013,Heggie2014,Ivanova2017,Kremer2018a} and open clusters \citep[e.g.,][]{Banerjee2018b}. In this section, we discuss the various types of BH binaries that form in the cluster models of this analysis.

Figure \ref{fig:scatter} shows the BH mass, $M_{\rm{BH}}$, and semi-major axis, $a$, for all BH--LC binaries found in our models at late times. Here yellow and blue circles indicate systems with main sequence (MS) and white dwarf (WD) companions, respectively, while red stars represent sub-subgiant companions. Filled circles/stars represent systems in mass-transferring configurations and open circle/stars represent detached binaries.

Note that if a particular system is not broken by dynamical encounters, the same system may appear across multiple cluster shapshots. In Figure \ref{fig:scatter}, we simply show the orbital parameters of each system at the first time it appears in a late-time snapshot, so that the same binary does not appear multiple times in the figure.

In total, there are 29 distinct accreting BH--LC binaries found at snapshots in the range $t=10-12$ Gyr in all of our models. Of these, 26 have MS donors, 2 have WD donors, and 1 has a sub-subgiant donor. There are 68 total detached BH--LC binaries, including 50 with MS companions, 17, with WD companions, and 1 with a sub-subgiant companion.

Sub-subgiants (SSGs), which are similar to the so-called ``red straggler" stars, occupy a unique location in the color-magnitude diagram where standard single-star evolution does not predict stars to exist. These stars lie redward of the normal MS stars but are fainter than the subgiant branch. SSGs have been observed and studied in several open and globular clusters \citep{Belloni1998,Albrow2001,Geller2017a,Shishkovsky2018}.

From a theoretical perspective, the formation of SSGs and/or red straggler stars has been explored at length in several analyses \citep[e.g.,][]{Leiner2017,Geller2017c,Ivanova2017}. In particular, \citet{Geller2017c} performed a detailed analysis showing that SSGs can form through several distinct channels in GCs. Using the SSG formation channels outlined in \citet{Geller2017c} (see Section 5.2 of that paper, which studies SSG formation in cluster models also produced using \texttt{CMC}), we identify two BH--SSG binaries in our models: one formed through the ``SG Mag'' channel (open red star in Figure \ref{fig:scatter}; see Section 2.2 of \citet{Geller2017a}) and one through the ``MS Coll'' channel (filled red star in Figure \ref{fig:scatter}; see Section 2.4 of \citet{Geller2017c}).

The variety of companion stellar types identified in our models in Figure \ref{fig:scatter} is in line with observations of companions to BH candidates observed to date in GCs. The BH-candidate in NGC 3201 (observed in a detached configuration) is identified to have a MS companion with mass near the turnoff mass. For two of the observed accreting BH candidates found to-date in GCs, WD companions are determined to be either plausible \citep[M22;][]{Strader2012} or confirmed \citep[e.g.,][]{Bahramian2017}. Finally, observations of the BH-candidate in M10 suggest the binary companion may be a SSG/red-straggler star. As shown in Figure \ref{fig:scatter}, all three of these companion types are produced naturally in our models.

Like the detached BH--MS binaries considered in \citet{Kremer2018d}, all of these BH--LC systems are dynamically assembled through binary-mediated exchange encounters.

All MS and WD binaries shown in Figure \ref{fig:scatter} as binaries formed through what we \citep[and other analyses, e.g.,][]{Ivanova2010} define as the exchange encounter channel, meaning these binaries are dynamically assembled from binary-mediated exchange encounters and then hardened to the point of Roche lobe overflow by a combination of subsequent (non-exchange) encounters and the effects of binary evolution (e.g., tidal effects and, for double degenerate systems, general relativistic effects). See \citet{Kremer2018a} for a detailed discussion of the formation of binaries through exchange encounters.

As discussed in, e.g., \citet{Kremer2018a}, \citet{Ivanova2010}, and \citet{Naoz2016}, if a hierchichal triple system with a BH--LC inner binary is dynamically assembled (e.g., as the outcome of a binary--single or binary--binary resonant encounter), Lidov-Kozai oscillations may drive the inner binary to mass transfer. This so-called triple-mediated channel will also contribute to the total population of accreting BH binaries. However, as shown in \citet{Kremer2018a}, this channel is likely to increase the total number of accreting BH binaries by at most $\approx 10\%$, therefore we do not consider this channel here and simply note that the total number of accreting BH binaries shown in Figure \ref{fig:scatter} may increase modestly if the contribution of this triple-mediated channel is included.

\section{Conclusions and Discussion}
\label{sec:conclusion}

We have demonstrated that by exploring a small range in initial cluster size (parameterized in terms of the initial cluster virial radius) motivated by observations of young massive clusters, we can produce a spectrum of cluster types, ranging from core-collapsed to puffy. Furthermore, we have shown that the initial $r_v$ of a GC model has a substantial effect upon the total number of BHs retained in the cluster at late times. Within the range of virial radii considered in this study ($r_v=0.5-5$ pc), our models retain BH populations ranging in size from just a couple to over 600 BHs at $t=12$ Gyr.

In particular, we have shown that the set of 11 GC models computed for this study span the observational features of four different Milky Way GCs: NGC 3201, M10, M22, and NGC 6752, the former three of which contain observed stellar-mass BH candidates. By identifying those cluster models that most accurately match the observed surface brightness profiles and velocity dispersion profiles of specific Milky Way GCs, we predict the total number of BHs retained in these clusters. At the present-day, we predict M10 retains $39 \pm 9$ BHs, M22 retains $40 \pm 9$ BHs, and NGC 6752 retains $16 \pm 7$ BHs. As a follow-up to the results of \citet{Kremer2018d}, in which we used the magnitude of BH natal kicks to adjust the number of BHs retained at late times, we predict with our new set of models that NGC 3201 retains $121 \pm 10$ BHs. The numbers of BHs predicted here are consistent with the predictions of other recent analyses that have used alternative methods to constrain BH numbers \citep[e.g.,][]{Weatherford2017,Askar2018}. 

The post-birth retention fraction of BHs in this analysis is very different compared to that of \citet{Kremer2018d}. In the present analysis, dynamical encounters are the primary mechanism through which BHs are ejected (the retention fraction at birth is fixed between different models), as opposed to \citet{Kremer2018d}, where the BH retention is determined through a combination of dynamical interactions \textit{and} natal kicks, which, in that analysis, are varied between models. Nonetheless, we demonstrate a similar correlation to that shown in \citet{Kremer2018d} between present-day structural parameters and the total number of BHs retained. Additionally, for NGC 3201, we estimate a similar total number of BHs retained in the cluster at present. This indicates that for GCs that are sufficiently dynamically evolved, the specific way that BHs are removed may matter less in shaping the present-day structure of the host GC than the number of BHs retained at present.

Additionally, we have explored the formation of BH--LC binaries in our models through several possible formation channels. We demonstrated that BH binaries are readily produced in our models in both mass-transferring and detached configurations. Furthermore, these BH binaries are found with MS, WD, and SSG companions, in line with the companion types identified for observed GC BH candidates. 

Note that, as discussed in \citet{Shishkovsky2018}, the identification of the primary star in the M10 binary as a BH is uncertain. In fact, several other possible options are consistent with the observations. A neutron star--red straggler binary, a RS CVn, a WD--red straggler/SSG binary, and even an isolated red straggler/SSG may all be viable alternative options (see Section 4 of \citet{Shishkovsky2018} for further details). We note that we identify 73 WD--SSG binaries in our models at late times (using the channels described in \citet{Geller2017c} to identify SSGs), 23 of which are found in mass-transferring configurations. Thus, from a dynamical-formation perspective, a WD--SSG/red straggler may indeed be a viable explanation for the binary of interest in M10. As discussed in \citet{Shishkovsky2018}, follow-up observations of this binary are necessary to more precisely constrain the true nature of the system.

Additionally, we note that \citet{Ivanova2017} explored the formation of BH--red straggler binaries in GCs through the grazing tidal capture of giants by BHs and estimated a formation rate of $\sim 1$ BH--red straggler binary per 50 BHs per Gyr in a typical cluster. Treatment of this grazing capture process is beyond the scope of \texttt{CMC}, therefore, we neglect this channel here and simply note that consideration of this channel may lead to an increase in the number of BH--red straggler binaries. 

Indeed, many details pertaining to the formation and evolution of exotic stellar sources such as SSGs and red stragglers remain uncertain. More thorough study of the formation of such objects (and the ways these objects may interact with BH populations in GCs) is needed to more precisely constrain their nature.

As described in Section \ref{sec:method}, we fix all initial cluster parameters for our grid of models with the exception of the initial virial radius. This allows us to isolate the effect that the initial virial radius has upon the long-term retention of BHs and its effect upon the structural properties of the cluster at late times. However, in fixing other initial parameters, in particular the initial $N$ and Galactocentric distance, we must address several caveats.

First, although the clusters considered in this study all have approximately similar total cluster masses at the present day, they are not identical. In this case, fixing the initial $N$ in our models limits our ability to produce best-fit models that capture the range in total cluster mass of the GCs considered here. As discussed in Section \ref{sec:method}, by allowing for $10\%$ uncertainties in the heliocentric distances of each cluster, the SBPs for our models can be shifted slightly to the right or left, which effectively allows us to compensate for the limited range in total cluster mass of our fixed-$N$ grid.

For example, the published heliocentric distance of M22 is 3.2 kpc \citep{Harris1996}. To attain the SBP for our best-fit model, we adopt a value of $d=2.9$ kpc, which shifts the SBP slightly to the right relative to a choice of $d=3.2$ kpc. If the value of 3.2 kpc is assumed to be precise, this suggests that our models may be slightly undermassive relative to M22.

To test this, we also ran a single additional model with initial $N=10^6$ and $r_v=0.9$ pc, which, at late times, has a SBP and velocity dispersion profile that also effectively match M22, but for a heliocentric distance of $3.2$ kpc. The final mass of this model is $\sim 20 \%$ higher than in the best-fit M22 model discussed in Section \ref{sec:m22}. As expected, more BHs are produced initially in the $N=10^6$ model ($\sim 2000$ versus $\sim 1500$ in the $N=8\times 10^5$ model). As a result, the $N=10^6$ model retains 79 BHs at $t=12$ Gyr, slightly higher than the $N_{\rm{BH}}=49$ value given for our best-fit M22 model. We conclude that although a slightly more massive model may more accurately match the published heliocentric distance of M22, the number of BHs will likely not change significantly, at least within the uncertainties of this analysis.

Secondly, by fixing the initial Galactocentric distance of our models and by assuming circular orbits in the Galactic potential, we neglect possible close passages ($d < 8$ kpc) to the Galactic center which may arise from eccentric cluster orbits. For a summary of the dynamical evolution of GCs on eccentric orbits about the Galactic potential see, for example, \citet{Baumgardt2003}. As noted in that analysis, M10, M22, and NGC 6752 may all have eccentric cluster orbits, with estimated pericenter distances of 3.4, 2.9, and 4.8 kpc, respectively. However, as also noted in \citet{Baumgardt2003}, the dissolution timescales for these three clusters due to close passages near the Galactic center are all $\gtrsim 3$ Hubble times \citep[see Table 2 of][]{Baumgardt2003}, so our treatment of circular orbits at a fixed Galactocentric distance of 8 kpc is likely a reasonable approximation. 

Nonetheless, we note that more precise modeling of the GCs considered in this study may incorporate better constrained Galactocentric distances and cluster orbits in the Galactic potential attained from, for example, \textit{Gaia}.

In reality, the process through which GCs are formed is likely more complex than considered in this analysis. The first few to 10s of Myrs of cluster evolution likely feature various complex processes such as hierarchical mergers and residual gas expulsion. Indeed, such processes are hinted at from observations of several young massive clusters \citep[e.g.,][]{Kuhn2014,Gennaro2017}. In particular, residual gas expulsion could lead to significant cluster expansion at early times, attenuating the long-term dynamical processing of BHs and thus altering the structural features of the clusters at late times. However, several recent analyses \citep[e.g.,][]{Banerjee2017,Brinkmann2017,Banerjee2018} have shown that the early stages of formation of some clusters may feature a substantially more compact embedded phase of sub-pc length scale, comparable to the thickest molecular-cloud filaments, from which a substantial gas dispersal would result in sizes comparable to those of the initial configurations considered here and also of the observed gas-free young massive clusters, as demonstrated in Figure \ref{fig:rc_obs}. While such processes may be important, they are beyond the scope of this analysis. The assumption here is that such formation processes are absorbed into the definition of initial conditions of our models.

Finally, we note that one notable feature of the core-collapsed cluster NGC 6752 is the presence of five observed millisecond pulsars (MSPs), which display unusual locations and/or accelerations compared to other pulsars observed in GCs \citep{Damico2002}. This suggests the occurrence of uncommon dynamics in NGC 6752.

In particular, one of these pulsars, PSR A, is observed at a distance of 6.39 arcmin from the gravitational center of NGC 6752 \citep{Damico2002}, the largest radial offset for any GC MSP observed to date. \citet{Colpi2003} argues that a four body-scattering event involving a stellar-mass BH--BH binary may be able to provide sufficient energy to eject PSR A to its current position in the cluster. Although a detailed examination of the formation and dynamical evolution of MSPs is beyond the scope of this paper, we note that our model most-closely matching NGC 6752 does contain 0--3 stellar-mass BH binaries at late times, which are similar to those invoked in \citet{Colpi2003} to describe the peculiar location of PSR A. A more detailed study of the interaction between BHs and MSPs in GCs will be presented in a forthcoming study \citep{Ye2018}.



As discussed in \citet{Colpi2003}, although the presence of stellar-mass BH--BH binaries may be sufficient to explain the anomalously high position of PSR A in NGC 6752, it may be difficult for a stellar-mass BH binary to also explain the anomalously high accelerations of PSR B and PSR E. Instead, as discussed in, e.g., \citet{Ferraro2003} an intermediate-mass BH ($M\sim100-200\,M_{\odot}$) may be necessary. A more detailed study of the formation of intermediate-mass BHs and the role that such objects play in the evolution of their host cluster is beyond the scope of this paper, and we defer such analysis to a future study. 






\acknowledgments
We thank the anonymous referee for their helpful comments and suggestions. This work was supported by NASA ATP Grant NNX14AP92G 
and NSF Grant AST-1716762. K.K. acknowledges support by the National Science Foundation Graduate Research Fellowship Program under Grant No. DGE-1324585.
S.C. acknowledges support from
CIERA, the National Aeronautics and Space Administration
through a Chandra Award Number TM5-16004X/NAS8-
03060 issued by the Chandra X-ray Observatory Center
(operated by the Smithsonian Astrophysical Observatory for and on behalf of the National Aeronautics
Space Administration under contract NAS8-03060), 
and Hubble Space Telescope Archival research 
grant HST-AR-14555.001-A (from the Space Telescope 
Science Institute, which is operated by the Association of Universities for Research in Astronomy, Incorporated, under NASA contract NAS5-26555).

\listofchanges


\newpage

\begin{thebibliography}{}
\bibitem [Abbott et al.$\,$(2016a)]{Abbott2016a} Abbott, B. P., Abbott, R., Abbott, T. D., et al. 2016a, ApJL, 818, L22
\bibitem [Abbott et al.$\,$(2016b)]{Abbott2016b} Abbott, B. P., Abbott, R., Abbott, T. D., et al. 2016b, PhRvL, 116, 061102
\bibitem [Abbott et al.$\,$(2016c)]{Abbott2016c} Abbott, B. P., Abbott, R., Abbott, T. D., et al. 2016c, PhRvL, 116, 241103
\bibitem [Abbott et al.$\,$(2016d)]{Abbott2016d} Abbott, B. P., Abbott, R., Abbott, T. D., et al. 2016d, PhRvL, 116, 061102
\bibitem [Abbott et al.$\,$(2016e)]{Abbott2016e} Abbott, B. P., Abbott, R., Abbott, T. D., et al. 2016e, PhRvL 118 221101
\bibitem [Abbott et al.$\,$(2017)]{Abbott2017}  Abbott, B. P., Abbott, R., Abbott, T. D., et al., PhRvL 118, 221101
\bibitem [Albrow et al.$\,$(2001)] {Albrow2001} Albrow M. D., Gilliland R. L., Brown T. M., Edmonds P. D., Guhathakurta P., \& Sarajedini A., 2001, ApJ, 559, 1060
\bibitem [Amaro-Seoane \& Chen et al.$\,$(2016)]{AmaroSeoane2016} Amaro-Seoane, P. \& Chen, X. 2016, MNRAS, 458, 3075
\bibitem [Antognini et al.$\,$(2014)] {Antognini2014} Antognini, J. M., Shappee, B. J., Thompson, T. A., \& Amaro-Seoane, P. 2014, MNRAS, 439, 1079
\bibitem [Askar et al.$\,$(2017)] {Askar2017} Askar, A., Szkudlarek, M., Gondek-Rosi\`{n}ska, D., Giersz,
M., \& Bulik, T. 2017, MNRAS, 464, L36
\bibitem [Askar et al.$\,$(2018)] {Askar2018} Askar, A., Arca Sedda, M. \& Giersz, M, 2018, arXiv:1802.05284
\bibitem [Arca Sedda et al.$\,$(2018)] {ArcaSedda2018} Arca Sedda, M., Askar, A., \& Giersz, M. 2018, arXiv:1801.00795
\bibitem [Bahramian et al.\,(2017)] {Bahramian2017} Bahramian, A., Heinke, C. O., Tudor, V., et al. 2017, MNRAS, 467, 2199
\bibitem [Banerjee et al.$\,$(2010)] {Banerjee2010} Banerjee, S. and Baumgardt, H. \& Kroupa, P. 2010, MNRAS, 402, 371
\bibitem [Banerjee \& Kroupa $\,$(2017)] {Banerjee2017} Banerjee, S. \& Kroupa, P., 2017, A\&A, 597, A28
\bibitem [Banerjee \& Kroupa $\,$(2018)] {Banerjee2018} Banerjee, S. \& Kroupa, P., 2018, ASSL, 424, 143
\bibitem [Banerjee$\,$(2018)] {Banerjee2018b} Banerjee, S., 2018, MNRAS, 481, 5123
\bibitem [Bastian et al.$\,$(2005)] {Bastian2005} Bastian, N., Gieles, M., Lamers, H. J. G. L. M., Scheepmaker, R. A., \& de Grijs, R. 2005, A\&A, 431, 905
\bibitem [Baumgardt et al.$\,$(2003)] {Baumgardt2003} Baumgardt, H. \& Makino, J. MNRAS, 430, 1
\bibitem [Belczynski et al.\,(2002)] {Belczynski2002} Belczynski, K., Kalogera, V., \& Bulik, T. 2002, ApJ, 572, 407
\bibitem [Belczynski et al.\,(2010)] {Belczynski2010} Belczynski, K., Bulik, T., Fryer, C. L., Ruiter, A., Valsecchi, F., Vink, J. S., \& Hurley, J. R. 2010a, ApJ, 714, 1217
\bibitem [Belloni et al.$\,$(1998)] {Belloni1998} Belloni T., Verbunt F., \& Mathieu R. D., 1998, A\&A, 339, 431
\bibitem [Breivik et al.$\,$(2016)] {Breivik2016} Breivik, K., Rodriguez, C. L., Larson, S. L., Kalogera, V., \& Rasio, F. A. 2018, ApJ, 830, L18
\bibitem [Brinkmann et al.$\,$(2017)] {Brinkmann2017} Brinkmann, N., Banerjee, S., Motwani, B., Kroupa, P. 2017, A\&A, 600, A49
\bibitem [Casterano \& Hut$\,$(1985)] {Casterano1985} Casertano, S., \& Hut, P. 1985, ApJ, 298, 80
\bibitem [Chatterjee et al.$\,$(2010)] {Chatterjee2010} Chatterjee, S., Fregeau, J. M., Umbreit, S., \& Rasio, F. A. 2010, ApJ, 719, 915
\bibitem [Chatterjee et al.$\,$(2013a)] {Chatterjee2013a} Chatterjee, S., Umbreit, S., Fregeau, J. M., \& Rasio, F. A. 2013a, MNRAS, 429, 2881
\bibitem [Chatterjee et al.$\,$(2013b)] {Chatterjee2013b} Chatterjee, S., Rasio, F. A., Sills, A., \& Glebbeek, E. 2013b, ApJ, 777, 106
\bibitem [Chatterjee et al.$\,$(2017a)] {Chatterjee2017a} Chatterjee, S., Rodriguez, C. L., \& Rasio, F. A. 2017a, ApJ, 834, 68
\bibitem [Chatterjee et al.$\,$(2017b)] {Chatterjee2017b} Chatterjee, S., Rodriguez, C. L., Kalogera, V., \& Rasio, F. A. 2017b, ApJ, 836, 26
\bibitem [Chomiuk et al.$\,$(2013)] {Chomiuk2013} Chomiuk, L., Strader, J., Maccarone, T. J., Miller-Jones, J. C. A., Heinke, C., Noyola, E., Seth, A. C., \& Ransom, S. 2013, ApJ, 777, 69
\bibitem [Colpi et al.$\,$(2003)] {Colpi2003} Colpi, M., Mapelli, M., \& Possenti, A. 2003, ApJ, 599, 1260
\bibitem [C\^{o}t\'{e} et al.$\,$(1994)] {Cote1994} C\^{o}t\'{e}, P., Welch, D. L., Fischer, P., Da Costa, G. S., Tamblyn, P et al. 1994, ApJSS, 90, 83
\bibitem [C\^{o}t\'{e} et al.$\,$(1995)] {Cote1995} C\^{o}t\'{e}, P., Welch, D. L., Fischer, P., \& Gebhardt, K. 1995, ApJ, 454, 788
\bibitem [Covino \& Ortolani$\,$(1997)] {Covino1997} Covino, S. \& Ortolani, S. 1997, A\&A, 318, 40
\bibitem [D'Amico et al.$\,$(2002)] {Damico2002} D'Amico, N., Possenti, A., Fici, L., Manchester, R. N., Lyne, A. G. et al. 2002, ApJ, 570, L89
\bibitem [Fall et al.$\,$(2005)]{Fall2005} Fall, S. M., Chandar, R. \& Whitmore, B. C. 2005, ApJ, 631, L133
\bibitem [Ferraro et al.$\,$(2003)] {Ferraro2003} Ferraro, F. R., Possenti, A., Sabbi, E., Lagani, P., Rood, R. T. et al. 2003, ApJ, 595, 179.
\bibitem [Forbes \& Bridges$\,$(2010)]{Forbes2010} Forbes, D. A. \& Bridges, T. 2010, MNRAS, 404, 3
\bibitem [Fregeau et al.$\,$(2003)] {Fregeau2003} Fregeau, J. M., Gu ̈rka, M. A., Joshi, K. J., \& Rasio, F. A. 2003, ApJ, 593, 772
\bibitem [Fregeau et al.$\,$(2004)] {Fregeau2004} Fregeau, J.M., Cheung, P., Portegies Zwart, S.F., \& Rasio, F. A. 2004,MNRAS, 352, 1 
\bibitem [Fregeau \& Rasio$\,$(2007)] {Fregeau2007} Fregeau, J. M., \& Rasio, F. A. 2007, ApJ, 658, 1047
\bibitem [Fryer \& Kalogera\,(2001)] {Fryer2001} Fryer, C. L. \& Kalogera, V. 2001, ApJ, 554, 548
\bibitem [Fryer et al.$\,$(2012)] {Fryer2012} Fryer, C. L., Belczynski K., Wiktorowicz G., Dominik M., Kalogera V., \& Holz D. E., 2012, ApJ, 749, 91
\bibitem [Geller et al.$\,$(2017a)] {Geller2017a} Geller, A. M., Leiner, E. M., Bellini, A., et al. 2017a, ApJ, 840, 66
\bibitem [Geller et al.$\,$(2017b)] {Geller2017c} Geller, A. M., Leiner, E. M., Chatterjee, S., Leigh, N. W. C., Mathieu, R. D., \& Sills, A. 2017b, ApJ, 842, 1
\bibitem [Gennaro et al.$\,$(2017)] {Gennaro2017} Gennaro, M., Goodwin, S. P., Parker, R. J., Allison, R. J., and Brandner, W. 2017, MNRAS, 472, 1760
\bibitem [Gieles et al.$\,$(2006)] {Gieles2006} Gieles, M., Larsen, S. S., Bastian, N., \& Stein, I. T. 2006, A\&A, 450, 129
\bibitem [Giesers et al.$\,$(2018)] {Giesers2018} Giesers, B., Dreizler, S., Husser, T.-O., Kamann, S., Escud\'{e}, G. A., et al. 2018, MNRAS, 475, L15
\bibitem [Harris$\,$1996$\,$(2010 edition)] {Harris1996} Harris, W.E. 1996, AJ, 112, 1487
\bibitem [Heggie \& Hut$\,$(2003)] {Heggie2003} Heggie D. \& Hut P. 2003, The Gravitational Million-Body Problem: A Multidisciplinary Approach to Star Cluster Dynamics. Cambridge University Press, Cambridge
\bibitem [Heggie \& Giersz$\,$(2014)] {Heggie2014} Heggie, D. C. \& Giersz, M. 2014, MNRAS, 439, 2459
\bibitem [H\'{e}non$\,$(1971a)] {Henon1971a} H\'{e}non, M. 1971a, AP\&SS, 14, 151
\bibitem [H\'{e}non$\,$(1971b)] {Henon1971b} H\'{e}non, M. 1971b, AP\&SS, 13, 284
\bibitem [Hobbs et al.$\,$(2005)] {Hobbs2005} Hobbs, G., Lorimer, D. R., Lyne, A. G., Kramer, M. 2005, MRNAS, 360, 974
\bibitem [Holtzman et al.$\,$(1992)] {Holtzman1992} Holtzman, J. A. et al. 1992, AJ, 103, 691
\bibitem [Hurley et al.$\,$(2000)] {Hurley2000} Hurley, J. R., Pols, O. R., \& Tout, C. A. 2000, MNRAS, 315, 543
\bibitem [Hurley et al.$\,$(2002)] {Hurley2002} Hurley, J. R., Tout, C. A. \& Pols, O. R. 2002, MNRAS, 329, 897
\bibitem [Hurley$\,$(2007)] {Hurley2007} Hurley, J. R. 2007, MNRAS, 379, 93
\bibitem [Hurley et al.$\,$(2016)] {Hurley2016} Hurley, J. R., Sippel, A. C., Tout, C. A., \& Aarseth, S. J.
2016, PASA, 33, e036
\bibitem [Irwin et al.$\,$(2010)] {Irwin2010} Irwin, J. A., Brink, T. G., Bregman, J. N., \& Roberts, T. P. 2010, ApJL, 712, L1
\bibitem [Ivanova et al.$\,$(2010)] {Ivanova2010} Ivanova, N., Chaichenets, S., Fregeau, J., et al. 2010, ApJ, 717, 948
\bibitem [Ivanova et al.$\,$(2005)] {Ivanova2005} Ivanova, N., Rasio, F. A., Lombardi, J. C., Jr., Dooley, K. L., \& Proulx, Z. F.
2005, ApJL, 621, L109
\bibitem [Ivanova et al.$\,$(2017)] {Ivanova2017} Ivanova, N.da Rocha, C. A., Van, K. X., \& Nandez, J. L. A. 2017, ApJL, 843, L30.
\bibitem [Joshi et al.$\,$(2000)] {Joshi2000} Joshi, K. J., Rasio, F. A., \& Portegies Zwart, S. 2000, ApJ, 540, 969
\bibitem [Joshi et al.$\,$(2001)] {Joshi2001} Joshi, K. J., Nave, C. P., \& Rasio, F. A. 2001, ApJ, 550, 691
\bibitem [Kamann et al.$\,$(2018)] {Kamann2017} Kamann, S., Husser, T.-O., Dreizler, S., Emsellem, E., Weilbacher, P. M. et al. 2018, MNRAS, 473, 4
\bibitem [King$\,$(1962)] {King1962} King, I. 1962, AJ, 67, 471
\bibitem [Kremer et al.$\,$(2018a)] {Kremer2018a} Kremer, K., Chatterjee, S., Rodriguez, C. L., \& Rasio, F. A. 2018a, ApJ, 852, 29
\bibitem [Kremer et al.$\,$(2018b)] {Kremer2018d} Kremer, K., Ye, C. S., Chatterjee, S., Rodriguez, C. L., \& Rasio, F. A. 2018b, ApJ, 855, L15
\bibitem [Kremer et al.$\,$(2018c)] {Kremer2018c} Kremer, K., Chatterjee, S., Breivik, K., Rodriguez, C. L., Larson, S. L., \& Rasio, F. A. 2018c, PRL, 120, 191103
\bibitem [Kroupa$\,$(2001)] {Kroupa2001} Kroupa, P. 2001, MNRAS, 322, 23
\bibitem [Kuhn et al.$\,$(2014)] {Kuhn2014} Kuhn, M. A., Feigelson, E. D., Getman, K. V., et al., 2014, ApJ, 787, 107
\bibitem [Kulkarni et al.$\,$(1993)] {Kulkarni1993} Kulkarni, S. R., Hut, P., \& McMillan, S. 1993, Nature, 364, 421
\bibitem [Lombardi et al.$\,$(2006)] {Lombardi2006} Lombardi, J. C., Proulx, Z. F., Dooley, K. L., Theriault, E. M., Ivanova, N., \& Rasio, F. A. ApJ,  640, 441
\bibitem [Leiner et al.$\,$(2017)] {Leiner2017} Leiner, E., Mathieu, R. D., Geller, A. M. 2017, ApJ, 840, 67
\bibitem [Maccarone et al.$\,$(2007)] {Maccarone2007} Maccarone, T. J., Kundu, A., Zepf, S. E., \& Rhode, K. L. 2007, Nature, 445, 183
\bibitem [Mackey et al.$\,$(2007)] {Mackey2007} Mackey, A. D., Wilkinson, M. I., Davies, M. B., \& Gilmore, G. F. 2007, MNRAS, 379, L40
\bibitem [Mackey et al.$\,$(2008)] {Mackey2008} Mackey, A. D., Wilkinson, M. I., Davies, M. B., \& Gilmore, G. F. 2008, MNRAS, 386, 65
\bibitem [Mandel$\,$(2016)] {Mandel2016} Mandel, I. 2016, MNRAS, 456, 578
\bibitem [Merritt et al.$\,$(2004)] {Merritt2004} Merritt, D., Piatek, S., Portegies Zwart, S., \& Hemsendorf,
M. 2004, ApJL, 608, L25
\bibitem [McLaughlin et al.$\,$(2005)] {McLaughlin2005} McLaughlin, D. E. \& van der Marel, R. P. 2005, ApJS, 161, 304
\bibitem [Miller et al.$\,$(1997)] {Miller1997} Miller, B. W., Whitmore, B. C., Schweizer, F., Fall, S. M. 1997, AJ, 114, 2381
\bibitem [Miller-Jones et al.$\,$(2014)] {Miller-Jones2014} Miller-Jones, J., Maccarone, T., Chomiuk, L., Strader, J., Bogdanov, S., Sivakoff, G., \& Heinke, C. 2014a, A new black hole candidate in the globular cluster 47 Tucanae, ATNF Proposal
\bibitem [Moody \& Sigurdsson$\,$(2009)] {Moody2009} Moody, K., \& Sigurdsson, S. 2009, ApJ, 690, 1370
\bibitem [Morscher et al.$\,$(2013)] {Morscher2013} Morscher, M., Pattabiraman, B., Rodriguez, C., Rasio, F. A., \& Umbreit, S. 2015, ApJ, 800, 9
\bibitem [Morscher et al.$\,$(2015)] {Morscher2015} Morscher, M., Umbreit, S., Farr, W. M., \& Rasio, F. A. 2013, ApJL, 763, L15
\bibitem [Naoz et al.$\,$(2016)] {Naoz2016} Naoz, S., Fragos, T., Geller, A., Stephan, A. P., Rasio, F. A. 2016, ApJL 822, L24
\bibitem [Pattabiraman et al.$\,$(2013)] {Pattabiraman2013} Pattabiraman, B., Umbreit, S., Liao, W.-k., et al. 2013, ApJS, 204, 15
\bibitem [Portegies Zwart et al.$\,$(2010)] {PortegiesZwart2010} Portegies Zwart, S. F., McMillan, S. L. W., \& Gieles, M. 2010, ARA\&A, 48, 431
\bibitem [Pryor \& Meylan$\,$(1993)] {Pryor1993} Pryor C. \& Meylan G., 1993, in Djorgovski S. G., Meylan G., eds, Astronomical Society of the Pacific Conference Series Vol. 50, Structure and Dynamics of Globular Clusters, p. 357
\bibitem [Repetto et al.$\,$(2012)] {Repetto2012} Repetto, S., Davies, M. B., \& Sigurdsson, S. 2012, MNRAS, 425, 2799
\bibitem [Repetto et al.$\,$(2017)] {Repetto2017} Repetto S., Igoshev A. P., \& Nelemans G., 2017, MNRAS, 467, 298
\bibitem [Rodriguez et al.$\,$(2015)] {Rodriguez2015} Rodriguez, C. L., Morscher, M., Pattabiraman, B., et al. 2015, PhRvL, 115, 051101
\bibitem [Rodriguez et al.$\,$(2016)] {Rodriguez2016a} Rodriguez, C. L., Chatterjee, S., \& Rasio, F. A. 2016, PhRvD, 93, 84029
\bibitem [Rodriguez et al$\,$(2018)] {Rodriguez2018} Rodriguez, C. L., Amaro-Seoane, P., Chatterjee, S., \& Rasio, F. A. 2018, 120, 151101
\bibitem [Scheepmaker et al.$\,$(2007)] {Scheepmaker2007} Scheepmaker, R. A., Haas, M. R., Gieles, M., Bastian, N., Larson, S. S., \& Lamers, H. J. G. L. M. 2007, A\&A, 469, 925
\bibitem [Scheepmaker et al.$\,$(2009)] {Scheepmaker2009} Scheepmaker, R. A., Gieles, M., Haas, M. R., Bastian, N., \& Larson, S. S. 2009, in Richtler, T., Larsen, S.. eds, The Radii of Thousands of Star Clusters in M51 with HST/ACS. Springer, Heidelberg, p. 103
\bibitem [Sigurdsson \& Hernquist$\,$(1993)] {Sigurdsson1993} Sigurdsson, S. \& Hernquist, L. 1993, Nature, 364, 423
\bibitem [Sills et al.$\,$(2013)] {Sills2013} Sills, A., Glebbeek, E., Chatterjee, S., \& Rasio, F. A. 2013, ApJ, 777, 105
\bibitem [Sippel \& Hurley$\,$(2013)] {Sippel2013} Sippel, A. C. \& Hurley, J. R. 2013, MNRAS, 430, L30
\bibitem [Shishkovsky et al.$\,$(2018)] {Shishkovsky2018} Shishkovsky, L., Strader, J., Chomiuk, L., Bahramian, A., Tremou, E, et al. 2018, ApJ, 855, 55
\bibitem [Spitzer$\,$(1967)] {Spitzer1967} Spitzer, Jr., L. 1969, ApJL, 158, L139
\bibitem [Spitzer$\,$(1987)] {Spitzer1987} Spitzer L., 1987, Dynamical evolution of globular clusters. Princeton University Press, Princeton, NJp. 191
\bibitem [Strader et al.$\,$(2012)] {Strader2012} Strader, J., Chomiuk, L., Maccarone, T. J., Miller-Jones, J. C. A., \& Seth, A. C. 2012, Nature, 490, 71
\bibitem [Strader$\,$(2014)] {Strader2014} Strader, J. 2014, A black hole in the Galactic Globular Cluster, Chandra Proposal, M10
\bibitem[Trager et al.$\,$(1995)]{Trager1995} Trager, S.C., King, I.R., and Djorgovski, S. 1995, AJ 109, 218.
\bibitem[Umbreit et al.$\,$(2012)] {Umbreit2012} Umbreit, S., Fregeau, J. M., Chatterjee, S., \& Rasio, F. A. 2012, ApJ, 750, 31
\bibitem[Usher et al.$\,$(2017)]{Usher2017} Usher, C., Pastorello, N., Bellstedt, S., Alabi, A., Cerulo, P. et al. 2017, MNRAS, 468, 4
\bibitem [Vesperini \& Chernoff$\,$(1994)] {Vesperini1994} Vesperini E. \& Chernoff D. F. 1994, ApJ, 431, 231
\bibitem [Watkins et al.$\,$(2015)] {Watkins2015} Watkins, L. L., van der Marel, R. P., Bellini, A., \& Anderson, J. 2015, ApJ, 803, 29
\bibitem [Weatherford et al.$\,$(2018)] {Weatherford2017} Weatherford, N. C., Chatterjee, S., Rodriguez, C. L., \& Rasio, F. A. 2017, arXiv:1712.03979
\bibitem [Whitemore et al.$\,$(1995)] {Whitmore1995} Whitemore, B. C. \& Schweizer, F. 1995, AJ, 109, 960
\bibitem [Ye et al.$\,$(2018), in prep.] {Ye2018} Ye, C. S., et al. 2018, in preparation
\bibitem [Ziosi et al.$\,$(2014)] {Ziosi2014} Ziosi, B. M., Mapelli, M., Branchesi, M., \& Tormen, G. 2014, MNRAS, 441, 3703
\bibitem [Zocchi et al.$\,$(2012)] {Zocchi2012} Zocchi, A., Bertin, G., \& Varri, A. L. 2012, A\&A, 539, A65
\end{thebibliography}
\end{document}